# Analysis and Development of SiC MOSFET Boost Converter as Solar PV Pre-regulator

**A.Bharathi sankar [1], Dr.R.Seyezhai [2]**

*[1]Research scholar, Department of Electrical & Electronics Engineering, SSN college of Engineering Chennai, Tamilnadu, INDIA. bharathisankar.1987@gmail.com,*
*[2] Associate Professor, Department of Electrical & Electronics Engineering, SSN college of Engineering Chennai, Tamilnadu, INDIA. seyezhair@ssn.edu.in.*

**Abstract:** Renewable energy source such as photovoltaic (PV) cell generates power from the sun light by converting solar power to electrical power with no moving parts and less maintenance. A single photovoltaic cell produces voltage of low level. In order to boost up the voltage, a DC-DC boost converter is used. In order to use this DC-DC converter for high voltage and high frequency applications, Silicon Carbide (SiC) device is most preferred because of larger current carrying capability, higher voltage blocking capability, high operating temperature and less static and dynamic losses than the traditional silicon (Si) power switches. In the proposed work, the static and dynamic characteristics of SiC MOSFET for different temperatures are observed. A SiC MOSFET based boost converter is investigated which is powered by PV source. This DC - DC converter is controlled using a Pulse-Width Method (PWM) and the duty cycle *d* is calculated for tracking the maximum power point using incremental conductance algorithm of the PV systems implemented in FPGA. Simulation studies are carried out in MATLAB/SIMULINK.A prototype of the SiC converter is built and the results are verified experimentally. The performance parameters of the proposed converter such as output voltage ripple input current ripple and losses are computed and it is compared with the classical silicon (Si) MOSFET converter.

**Keywords:** Photovoltaic panel, Maximum power point tracking control, Silicon carbide, Incremental Conductance, Pulse Width Modulation, Field Programmable Gate Array.

## 1. Introduction:

For several decades, silicon (Si) has been the primary semiconductor choice for power electronic devices. However, Si is quickly approaching its limits in power conversion. Wide band gap (WBG) semiconductors offer improved efficiency, reduced size and lower system cost. Of the various types of WBG semiconductors, silicon carbide (SiC) have proven to be the most promising technologies, with several devices already being available commercially [1]. SiC has shown tremendous high temperature capability, as well as aptitude for high voltage applications. Furthermore, the cost of SiC devices has decreased within the last decade, and the performance has proven superior to that of conventional Si based devices. Some of the potential application areas for these devices include: transportation electrification and renewable energy. SiC devices have been explored for replacement of Si IGBTs in photovoltaic inverters in order to improve efficiency. Further, the fast switching speed of these devices allows for high frequency operation thereby resulting in the reduction of the passive components, which decreases the total size, weight and cost of the system [2].

The advantages of SiC have been investigated in this paper by developing a DC-DC boost converter using SiC MOSFET for PV applications. The characteristics of SiC MOSFET are analyzed and its performance is compared with the classical Si MOSFET. SiC based boost converter results in reduced input current and voltage ripple by implementing the PWM control using FPGA. Further, to get maximum power from PV, incremental conductance algorithm is employed and implemented using FPGA. Hardware setup of the proposed DC-DC converter is developed and the results are verified experimentally.

## 2. Description of SiC MOSFET

SiC MOSFETs being unipolar devices typically experience faster switching than an IGBT. As a result, extensive work on the characterization of SiC MOSFETs and comparison of their dynamic performance with Si IGBTs has been carried out. [3]. A common structure for SiC MOSFETs is the double-diffused or DMOSFET which is shown in Figure 1, that allows for fast switching speed and high durability. Upon the application of a positive gate bias, an inversion layer is produced at the surface of the p- well region underneath the gate electrode. This inversion layer provides a path for the flow of current from the drain to the source. This structure includes an intrinsic body diode, and allows operation both in the first and third quadrants. Current flows through the body diode when the MOSFET gate is off, and a positive drain bias exceeding approximately 0.7 V is applied. If instead a positive gate voltage is applied, and the drain is negatively biased, then the channel will conduct with current flowing from the source to the drain, resulting in third quadrant operation [4].

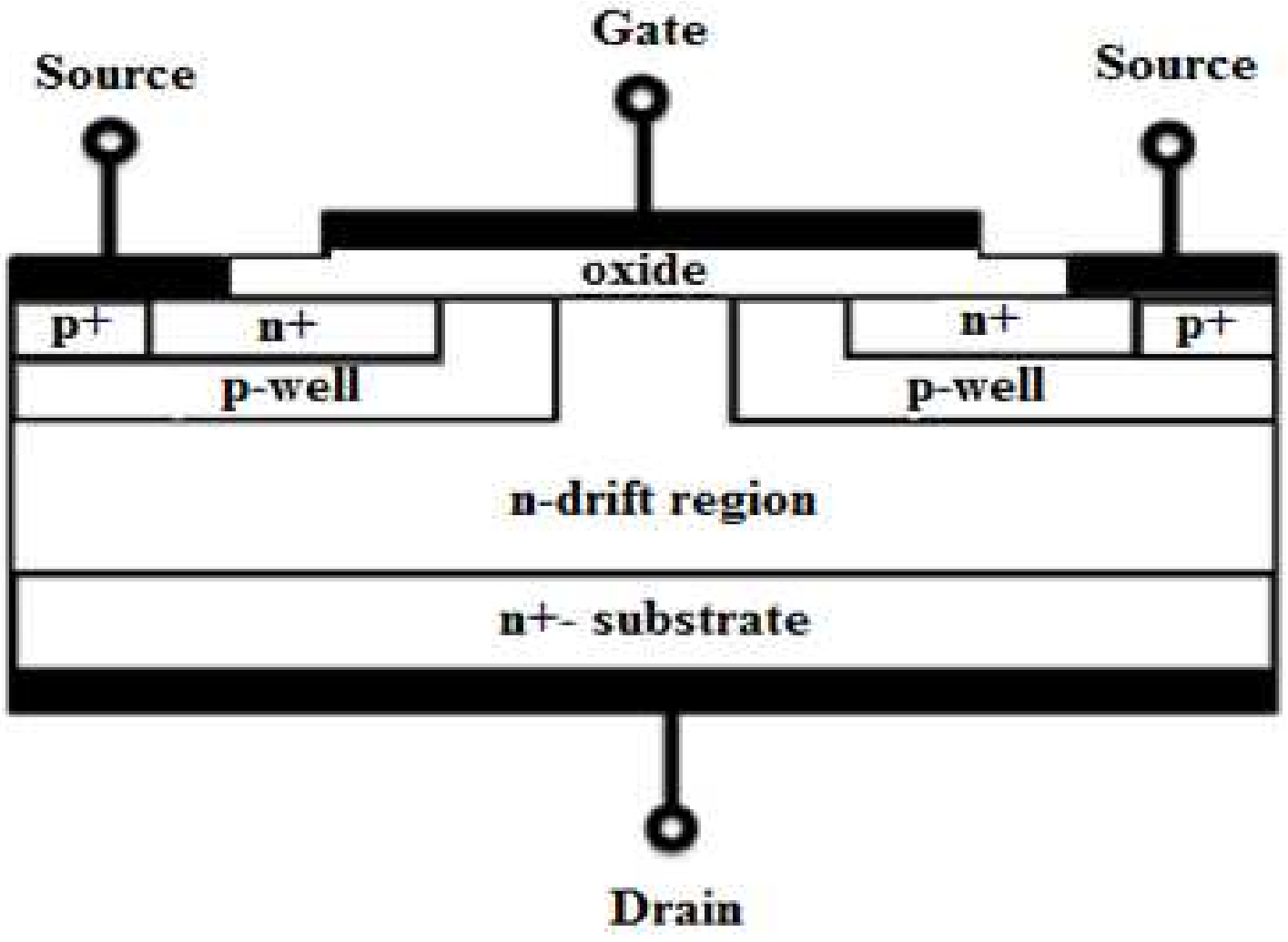

Figure 1: SiC MOSFET structure

SiC power MOSFET is also capable of supporting high positive drain voltages. Furthermore, due to its greater critical electric field for breakdown, the doping concentrations in the drift region of SiC MOSFETs can be increased, thereby resulting in a lower drift resistance for a given blocking voltage. This relationship is shown by the following equation for the ideal on-state resistance Ron-ideal

$$R_{on-ideal} = \frac{4BV^2}{\varepsilon_s \mu_n E_C^3} \quad \text{...........} \quad (1)$$

Where BV is the breakdown voltage, $\varepsilon_s$ is the dielectric constant of the semiconductor, $\mu_n$ is the mobility of the drift region, and Ec is the critical electric field for breakdown. Moreover, SiC also features a higher saturation drift velocity, allowing for faster switching, and thus is suitable for high frequency applications.

## 3. CHARACTERIZATION OF SiC MOSFET

SiC MOSFET is modeled in MATLAB and the static and dynamic characteristics are simulated by extracting the parameters from the data sheet for various temperatures.

### 3.1. Output Characteristics

Figure 2 shows the Simulink model for output characteristics for the SiC MOSFET. Output characteristic shows the variation of the drain current with respect to the drain source voltage at 25 °C and 150°C temperature as shown in figure 3 & 4. The output characteristics are Drain current $I_D$ versus Drain-source voltage $V_{DS}$ measured under different gate voltage VGS from 10 V to 22 V. It is found the results that the SiC MOSFET goes into saturation at prolonged period thereby pinch off point is increased.

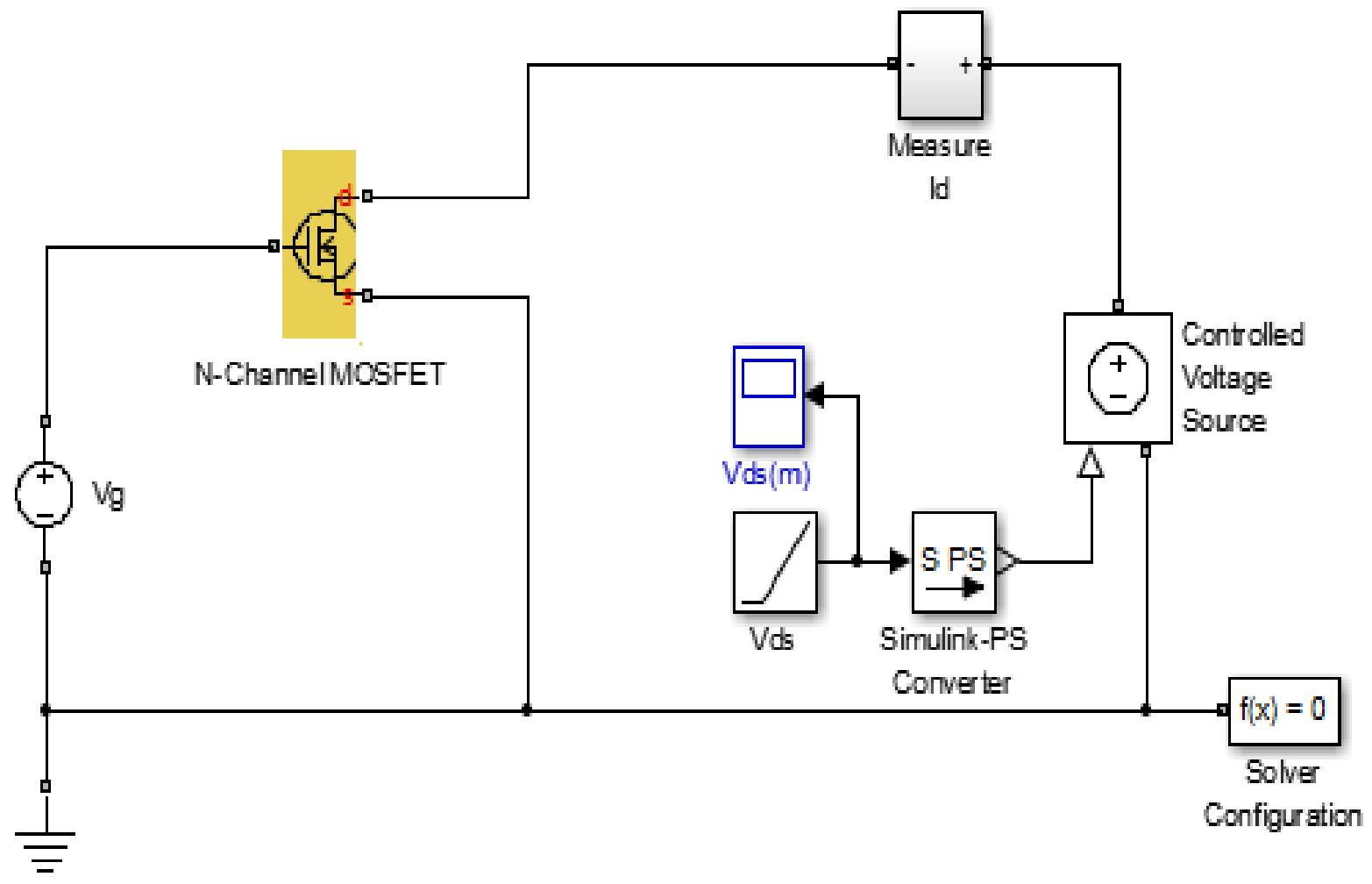

Figure 2: Simulink model of SiC MOSFET Static characteristics.

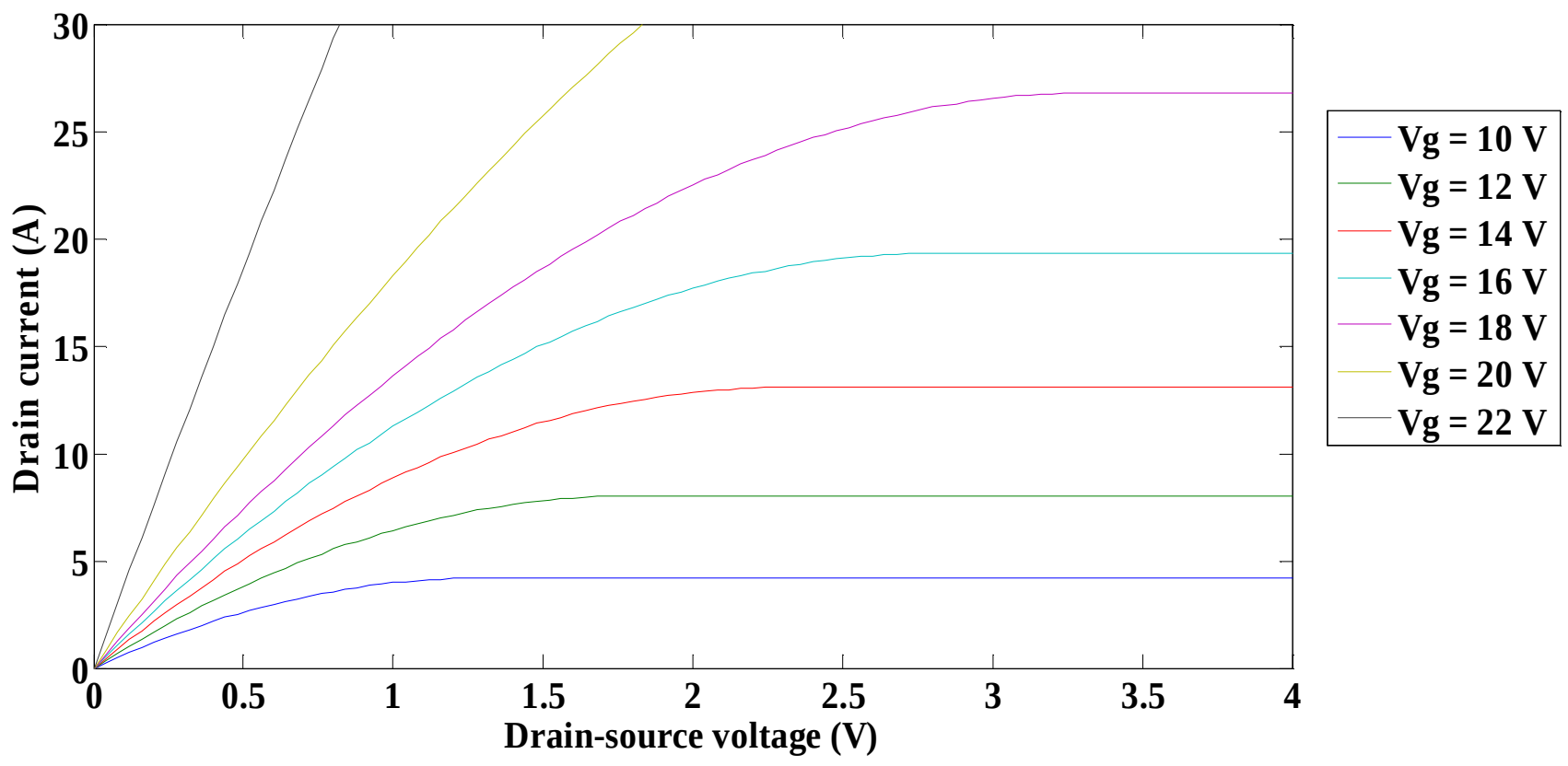

Figure 3: Output characteristics of SiC MOSFET at 25 C

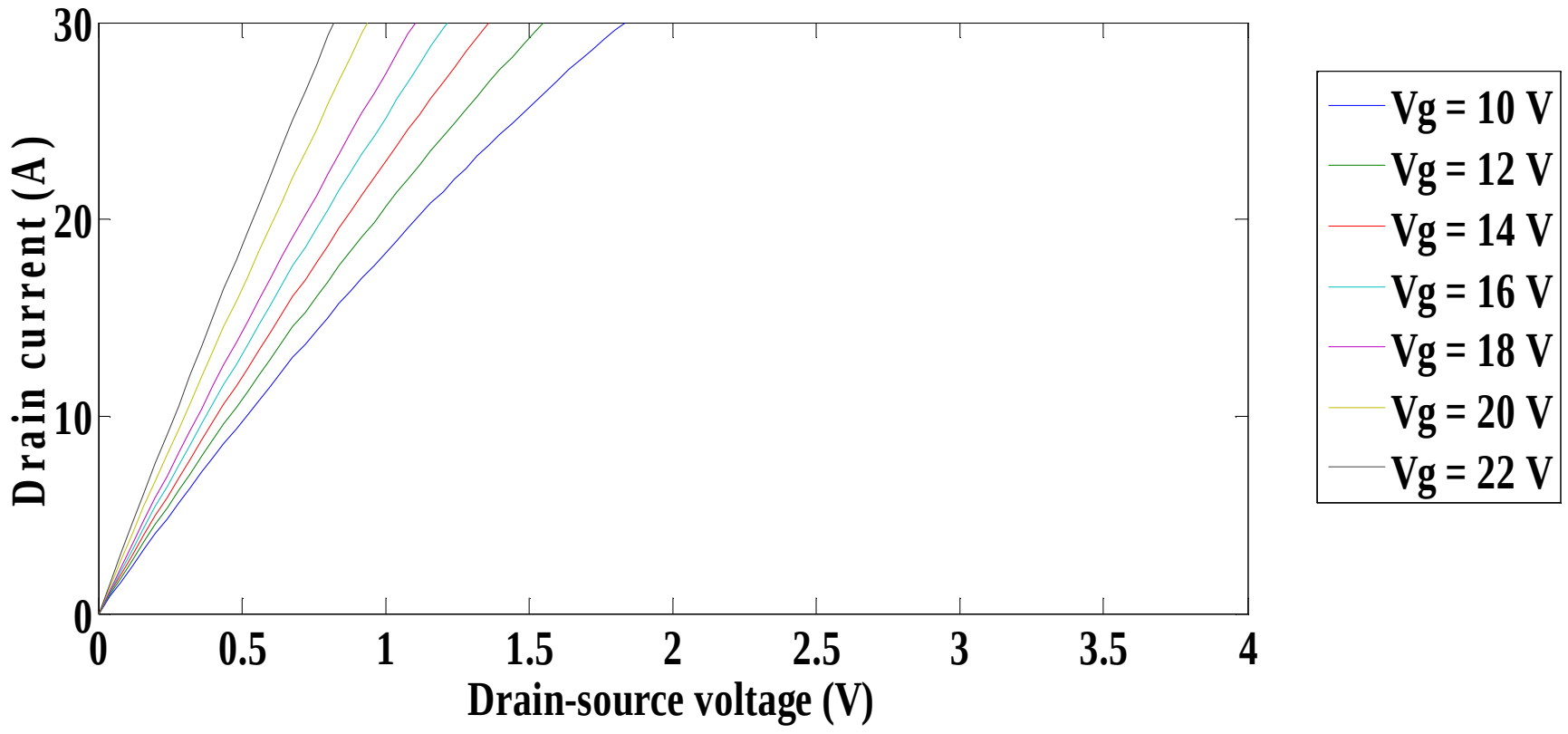

Figure 4: Output characteristics of SiC MOSFET at 150 C

### 3.2. Transfer Characteristics

Figure 5 shows the variation of the drain current with respect to the gate source voltage for various temperatures. From the figure it is clear that at high temperature, the threshold voltage of the device and transconductance increases.

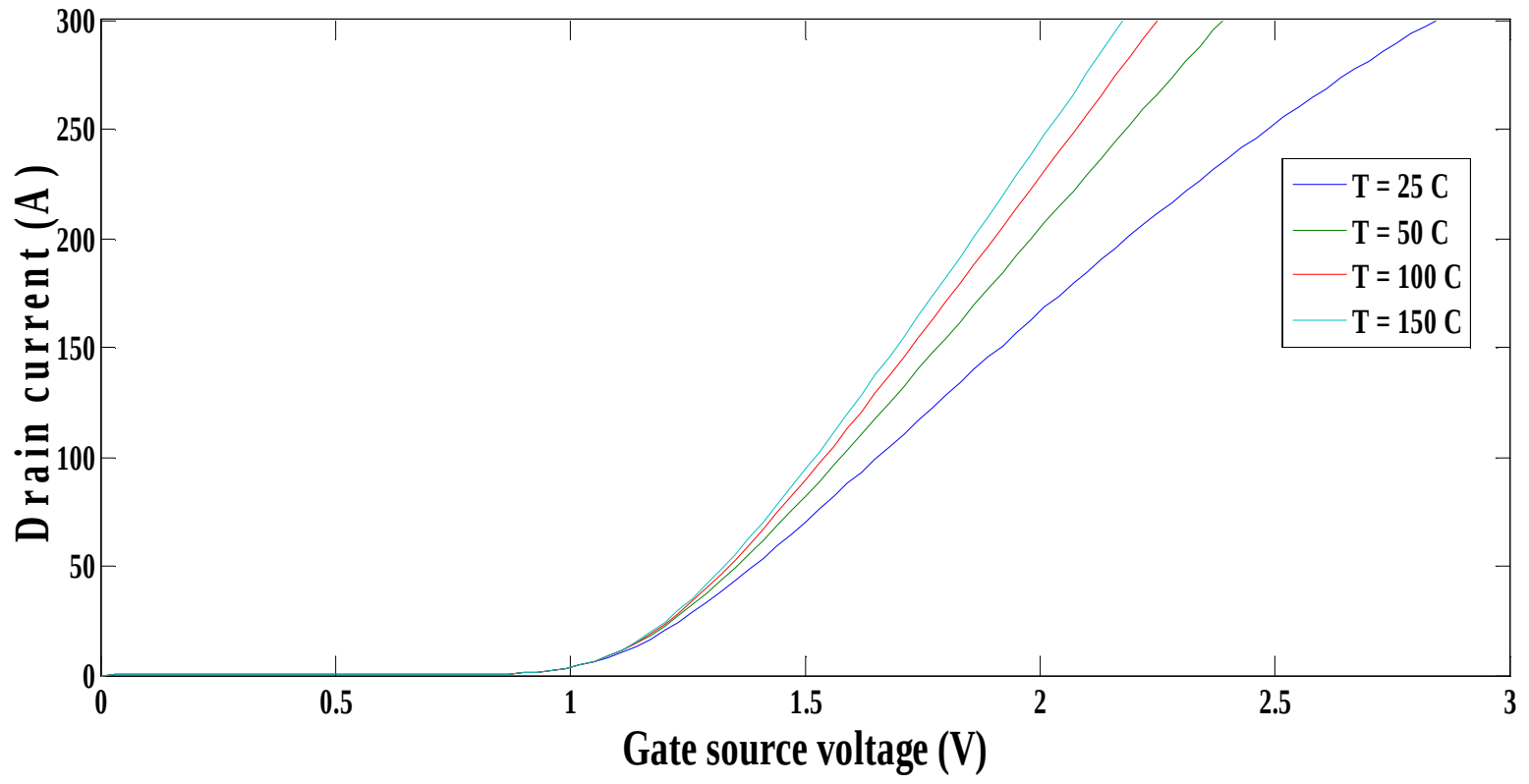

Figure 5: Transfer Characteristics of SiC MOSFET

### 3.3 On-State Resistance $R_{DS(on)}$

The on-state resistance $R_{DS(ON)}$ is a critical parameter to the device since it determines the conduction power dissipation. The power D-MOSFET structure with its eight internal resistance components between the drain and source electrodes when the device at turned-on state. The total on-state resistance is the sum of the eight resistances, which can be expressed as $R_{DS(ON)} = R_{CS} + R_{N}^{+} + R_{CH} + R_{A} + R_{JFET} + R_{D} + R_{SUB} + R_{CD}$. Where $R_{CS}$ is source contact resistance, $R_{N}+$ is the source resistance, $R_{CH}$ is channel resistance, $R_{A}$ is accumulation resistance, $R_{JFET}$ is JFET resistance, $R_{D}$ is drift region resistance, $R_{SUB}$ is the substrate resistance and $R_{CD}$ is the drain contact resistance. There are several different definitions for the $R_{DS(ON),}$ it to be the maximum slope of the output curve at a given turn-on gate voltage. This definition gives the minimum possible $R_{DS(ON)}$ for a given $V_{GS}$, which resulting in $R_{DS(ON)} = 0.129$ Ω at $V_{GS} = 20$ V in our case. In this work, $R_{DS(ON)}$ can be read directly from the output characteristic curves.

The $R_{DS}$ characteristic of Si and SiC MOSFET is shown in figure 5 respectively for various temperatures. It is clear that the variation of on state resistance is in milliohms for the entire range of temperature for SiC compared to Si.

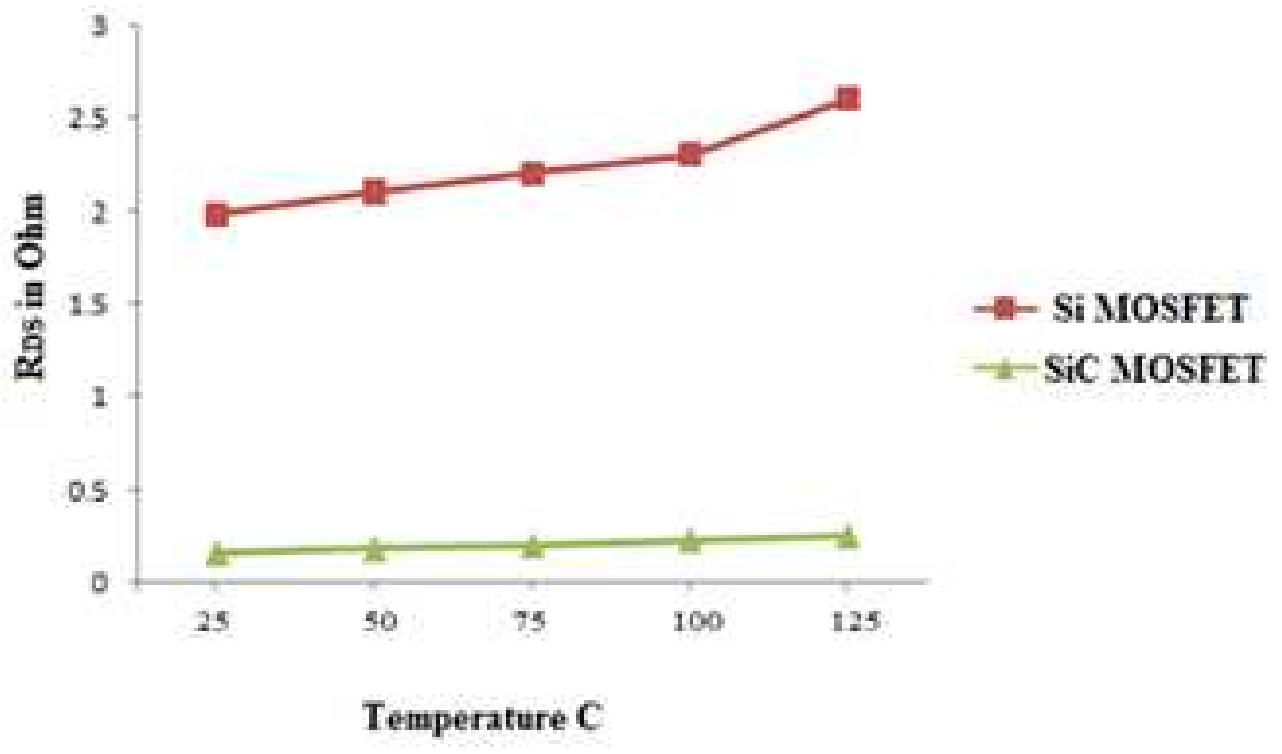

Figure 5: $R_{DS}$ characteristic of Si

### 3.3 Switching characteristics curve

Figure 6 shows the Simulink model for dynamic characteristics for the SiC MOSFET. Switching characteristics for SiC MOSFET is shown in figure 7 and it provides the information of the SiC MOSFET under transient and saturation region and its corresponding dynamic parameters Turn on time($t_{on}$)=35ns, Turn off time ($t_{off}$)=76ns, Fall time ($t_f$)=36ns & Rise time ($t_r$)=22ns is shown in fig 7a & 7b.Experimental setup Switching characteristics for SiC MOSFET is shown in figure 7c.

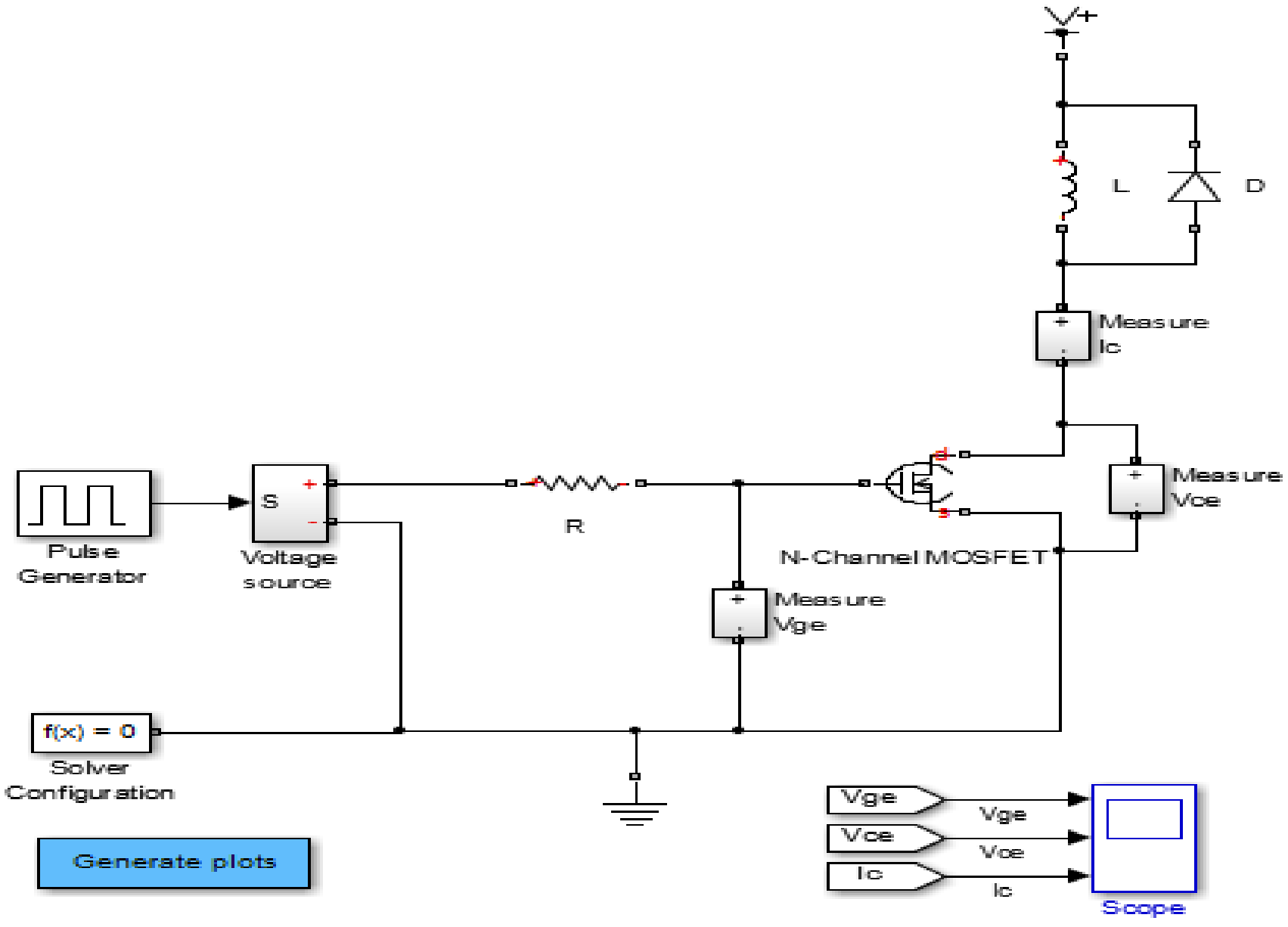

Figure 6: Simulink model of SiC MOSFET Dynamic characteristics.

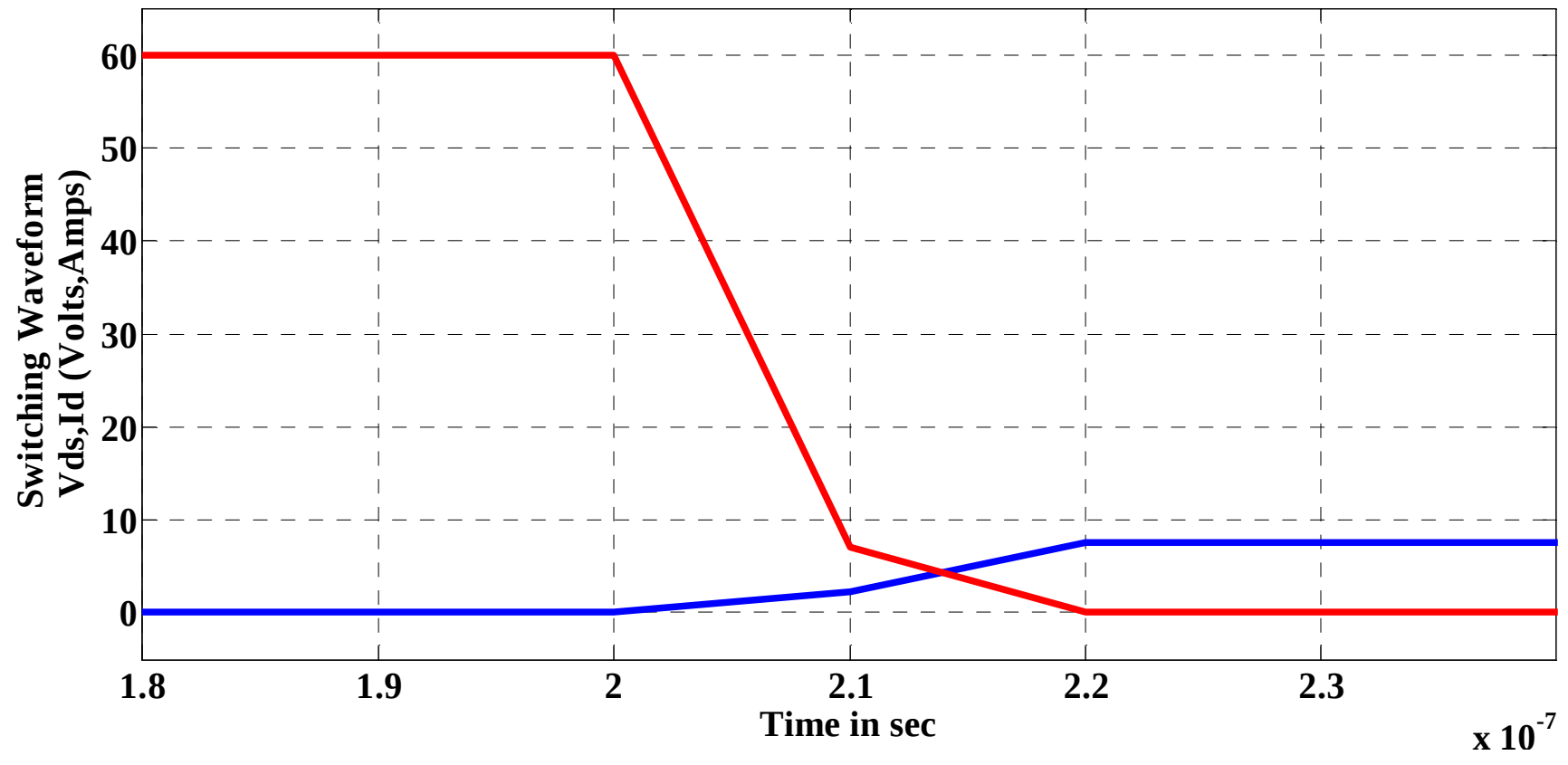

**Figure .7a** Switching characteristics $T_{on}$ state for SiC MOSFET

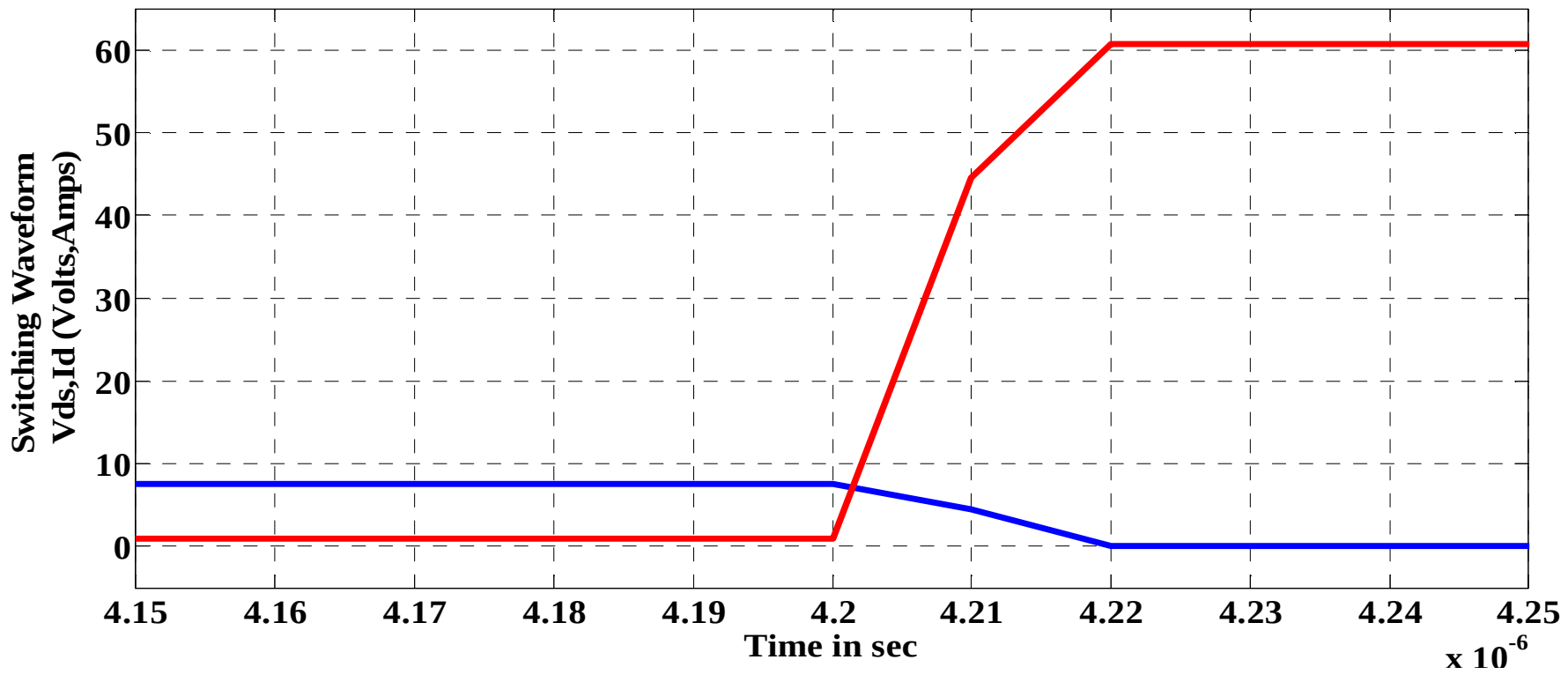

**Figure .7b** Switching characteristics $T_{off}$ state for SiC MOSFET

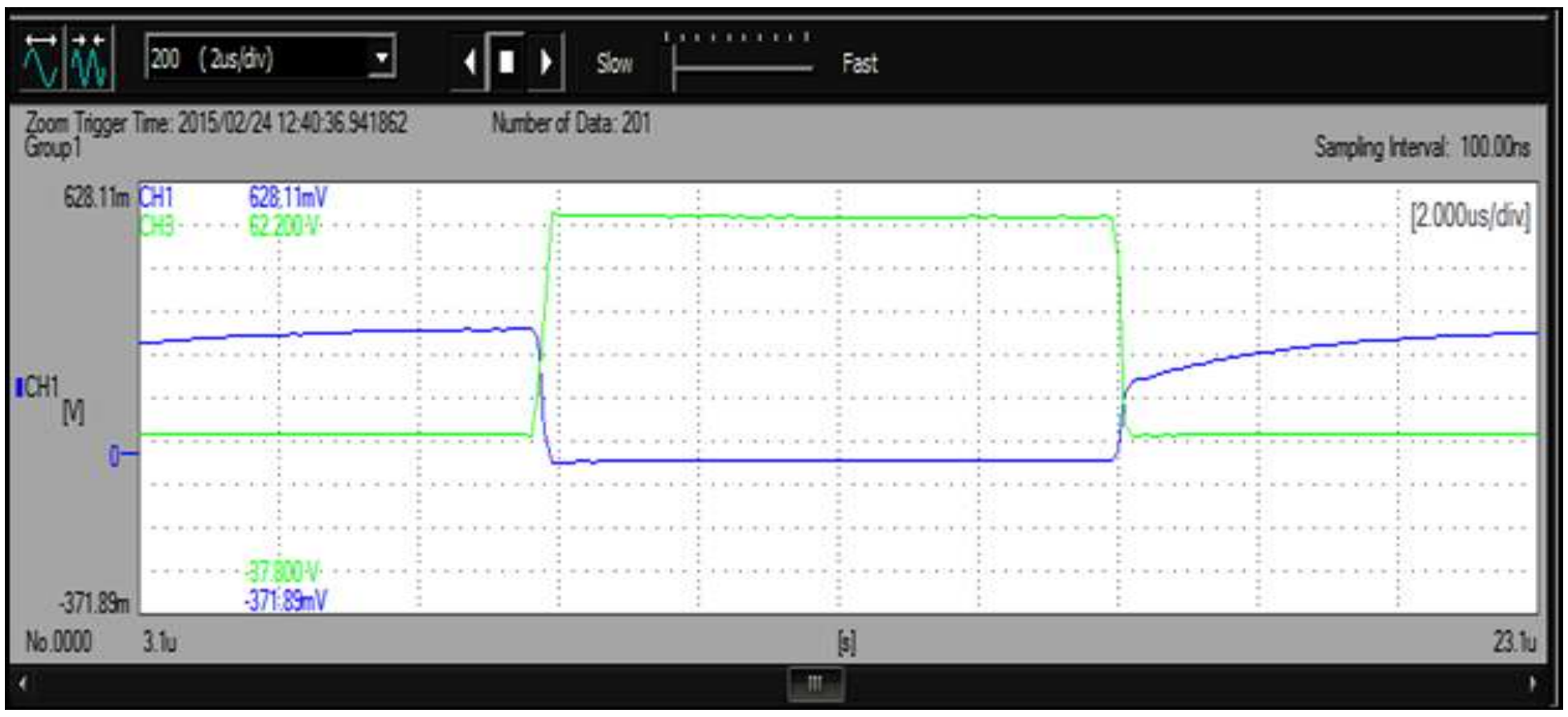

**Figure .7c** Experimental setup Switching characteristics for SiC MOSFET

### 3.4.Thermal Analysis for SiC MOSFET

As the power density and switching frequency increases, thermal analysis of power electronics system becomes imperative. The analysis provides valuable information on the semiconductor rating, long term reliability and efficient heat sink design have been reported in the literature for thermal analysis of SiC MOSFET. The aim of this work is to build a comprehensive thermal model for the SiC MOSFET modules. It is used in boost converter in order to predict the dynamic junction temperature rise under real operating conditions. The thermal model is developed in two steps, first step the losses are calculated and then the junction temperature is estimated. The real-time simulation environment dictates the requirement for the models for easy implementation on the software platform. The parameter of the thermal network is extracted from the junction to case and case to ambient dynamic thermal impedance curves. An equivalent RC network model is built to platform the thermal analysis as shown in Figure 8. It is shown in Figure 9 that the SiC MOSFET junction temperature and case temperature is about 400°C and 340°C respectively.

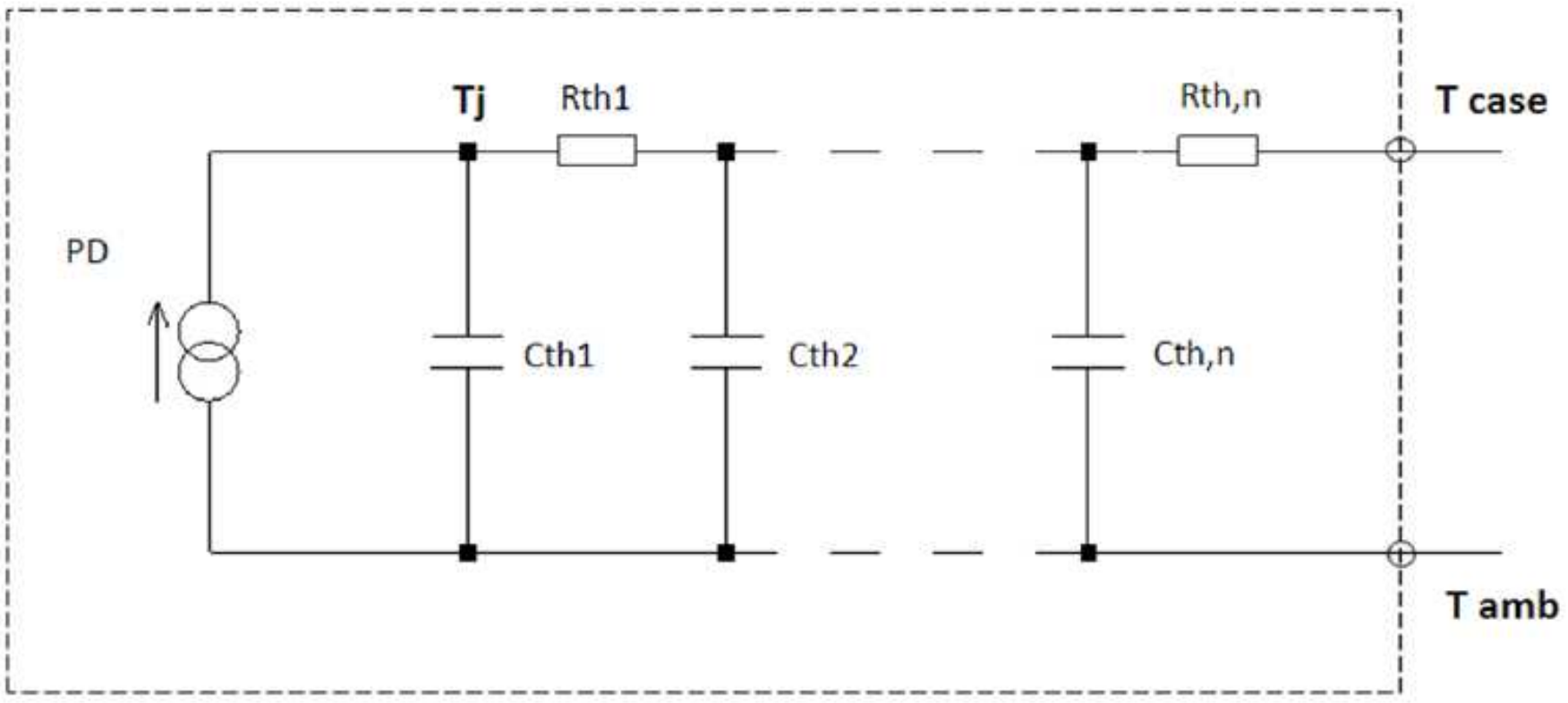

**Figure .8:** Thermal RC equivalent network model

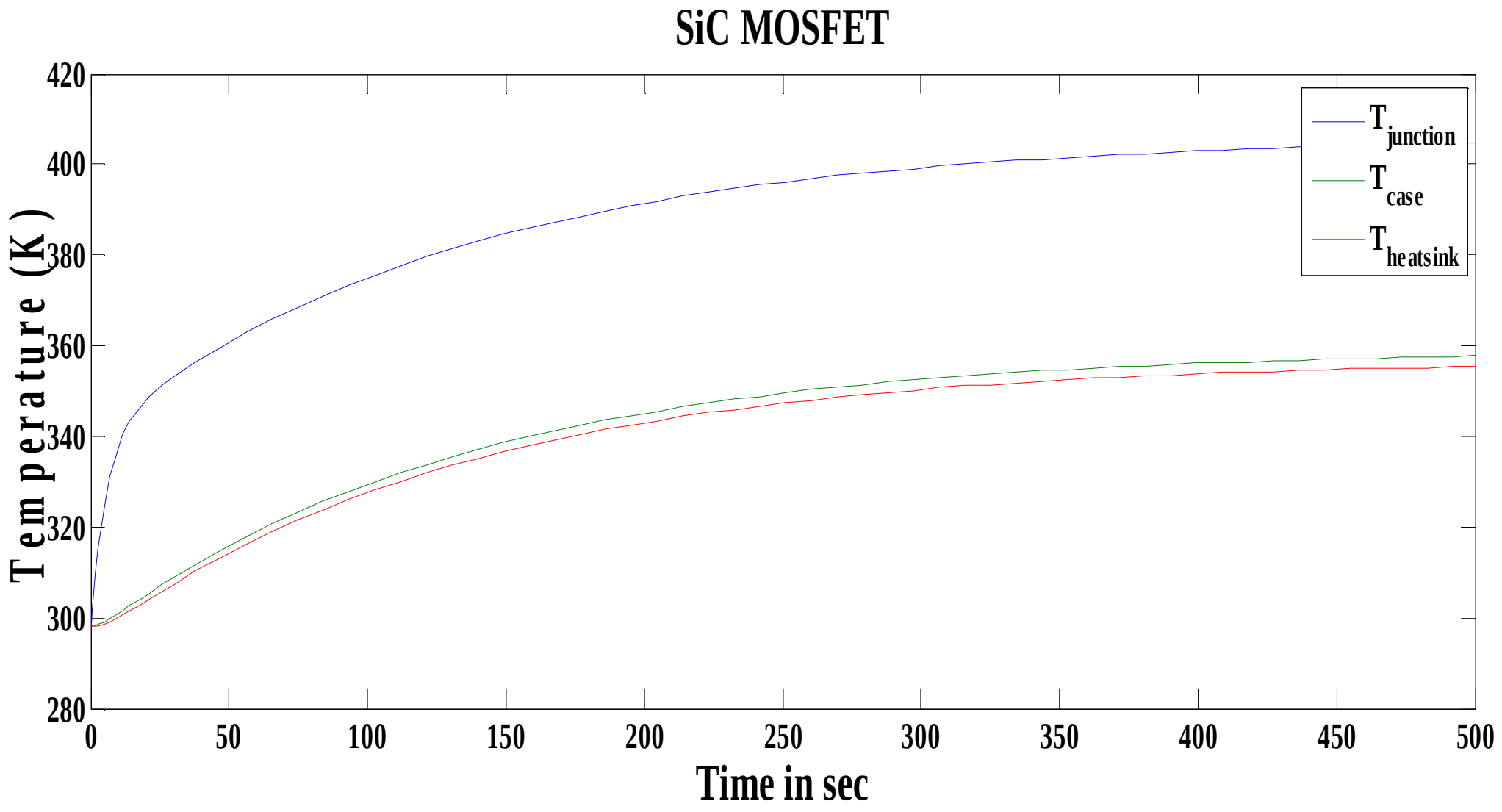

**Figure .9:** SiC MOSFET junction and case temperature

Figure 10 shows the Simulink model for Thermal characteristics for the SiC MOSFET. Thermal characteristic shows the variation of the drain current with respect to the drain source voltage at 25 °C and 125°C temperature as shown in figure 11.The output characteristics are Drain current $I_D$ versus Drain-source voltage $V_{DS}$ measured under different gate voltage VGS from 10 V to 22 V. It is found the results that the SiC MOSFET goes into saturation at prolonged period thereby pinch off point is increased.

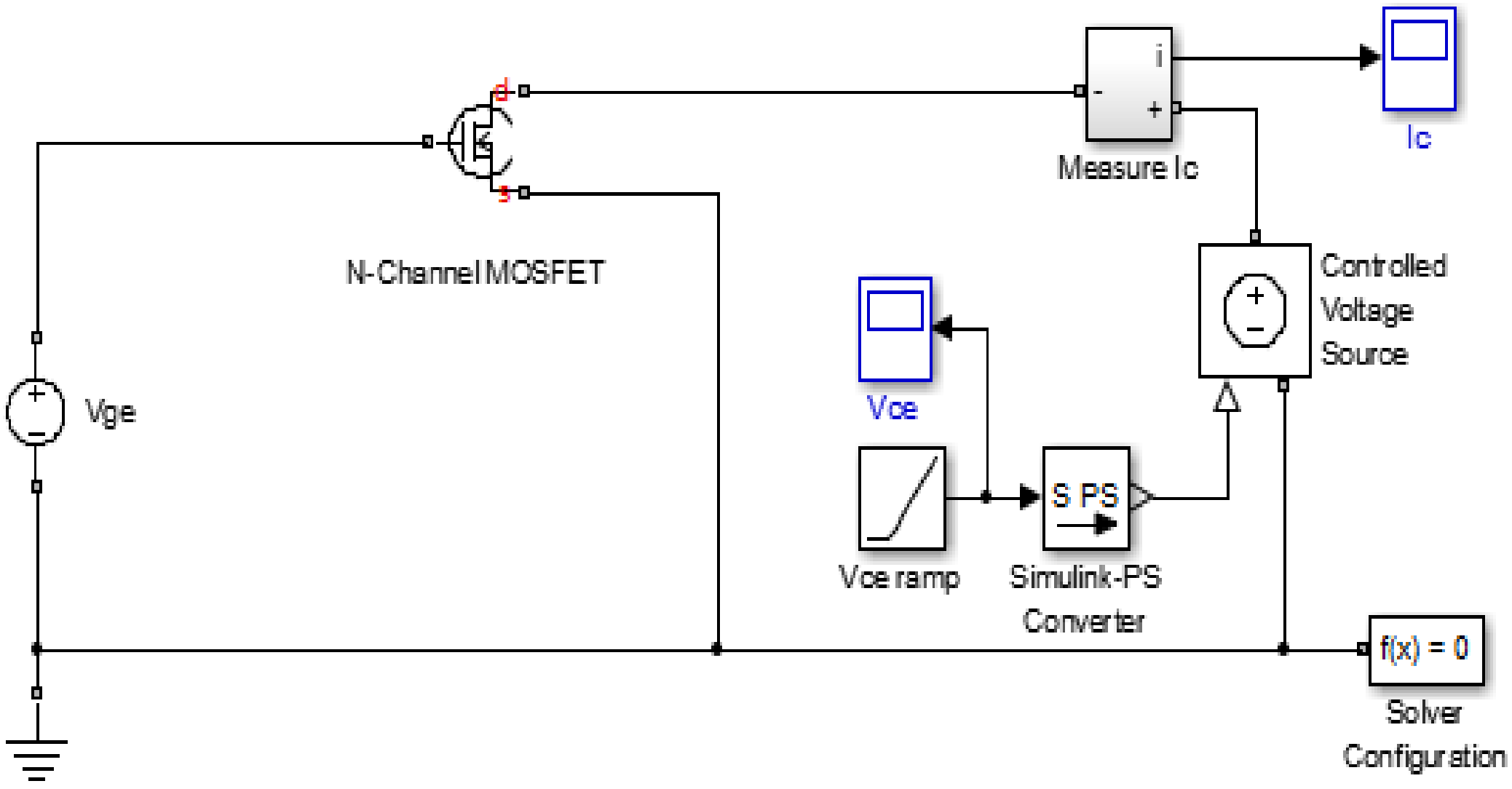

Figure 10: Simulink model of SiC MOSFET Thermal characteristics.

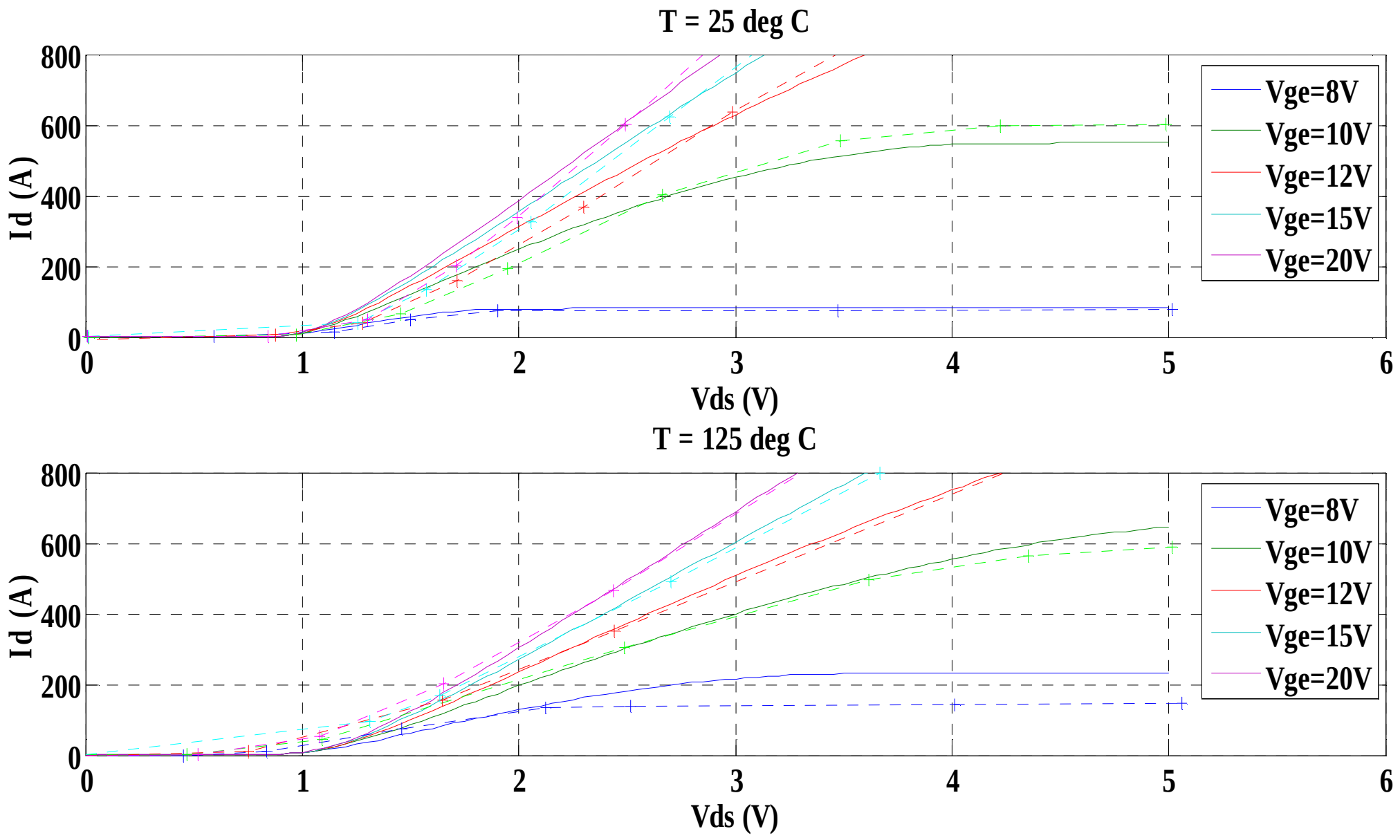

**Figure .11** Thermal characteristics for SiC MOSFET

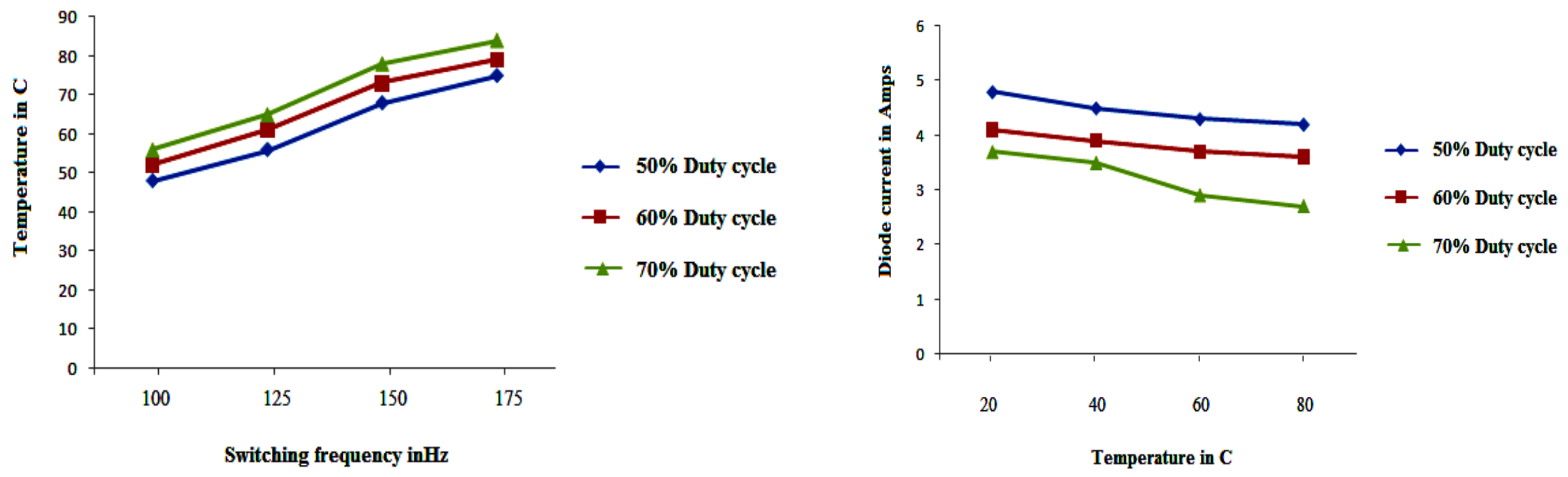

**Figure .11** Experimental setup temperature Vs switching frequency temperature Vs Drain current for SiC MOSFET for various duty cycle.

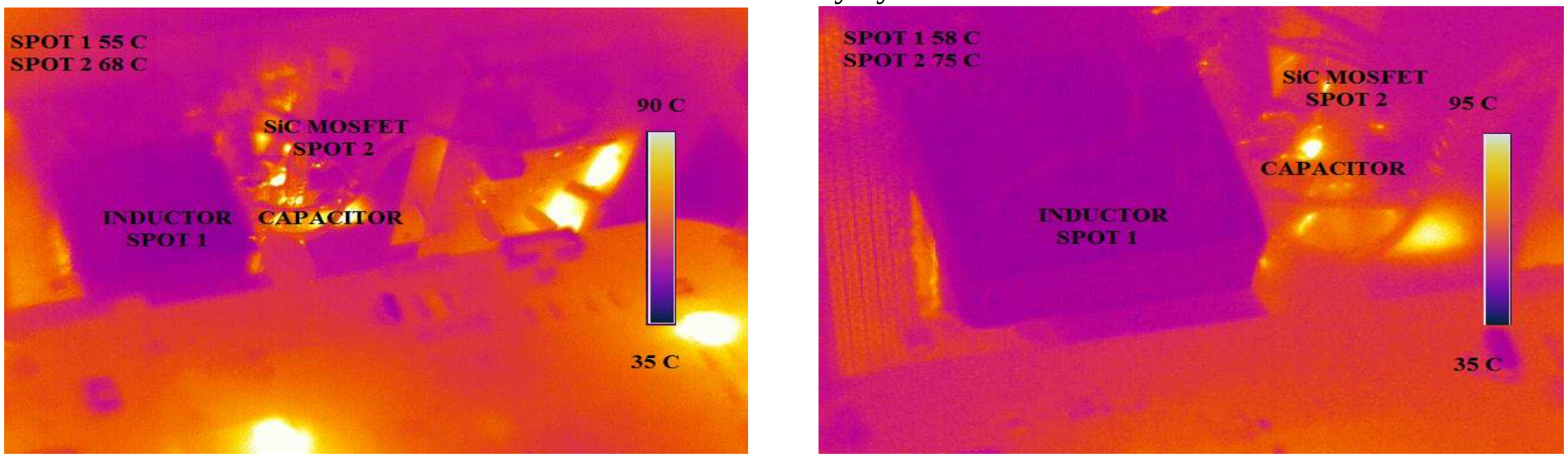

Fig.11. Thermal pictures of the main components in the boost converter for different switching frequencies

An infrared camera was used to investigate how the switching frequency affects the different components in the converter. For 100 kHz, 150 kHz and 175 kHz, temperatures of the diode, the SiC MOSFET and the inductor are shown in Fig.11. At a switching frequency of 150 kHz, the boost inductor has a temperature of 55 C. The heat sink, to which the SiC MOSFET is attached to, has a temperature of 68 C.Increasing the switching frequency to 175 kHz results in a high thermal stress in the SiC devices. The boost inductor has a temperature increases to 58 C. The case temperature of the SiC MOSFET increases to 75 C.

Experimental setup Thermal characteristic for SiC MOSFET is shown in figure 11.The SiC MOSFET junction temperature Vs switching frequency for various duty cycle and SiC MOSFET junction temperature Vs Drain current for various duty cycle.

## 4. **SiC MOSFET BASED DC-DC CONVERTER**

Choppers are static DC-DC converters for generating variable DC voltage source from a fixed DC voltage source. It is used to step up the input voltage to a required output voltage without the use of a transformer. The control strategy lies in the manipulation of the duty cycle of the switch which causes the voltage change. The circuit diagram of the designed SiC boost converter is shown in Figure.8.

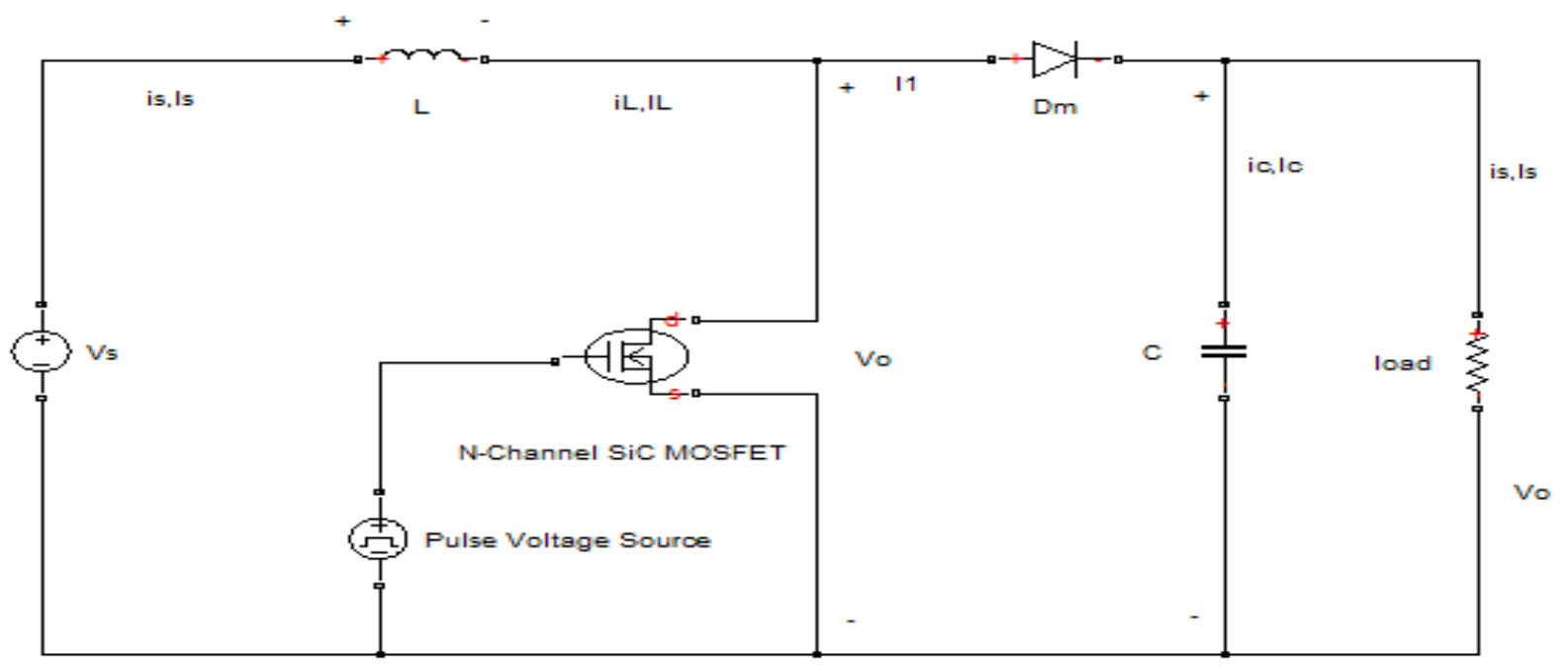

Figure 8: Circuit diagram of boost converter

The function of boost converter can be divided into two modes, Mode 1 and Mode 2.

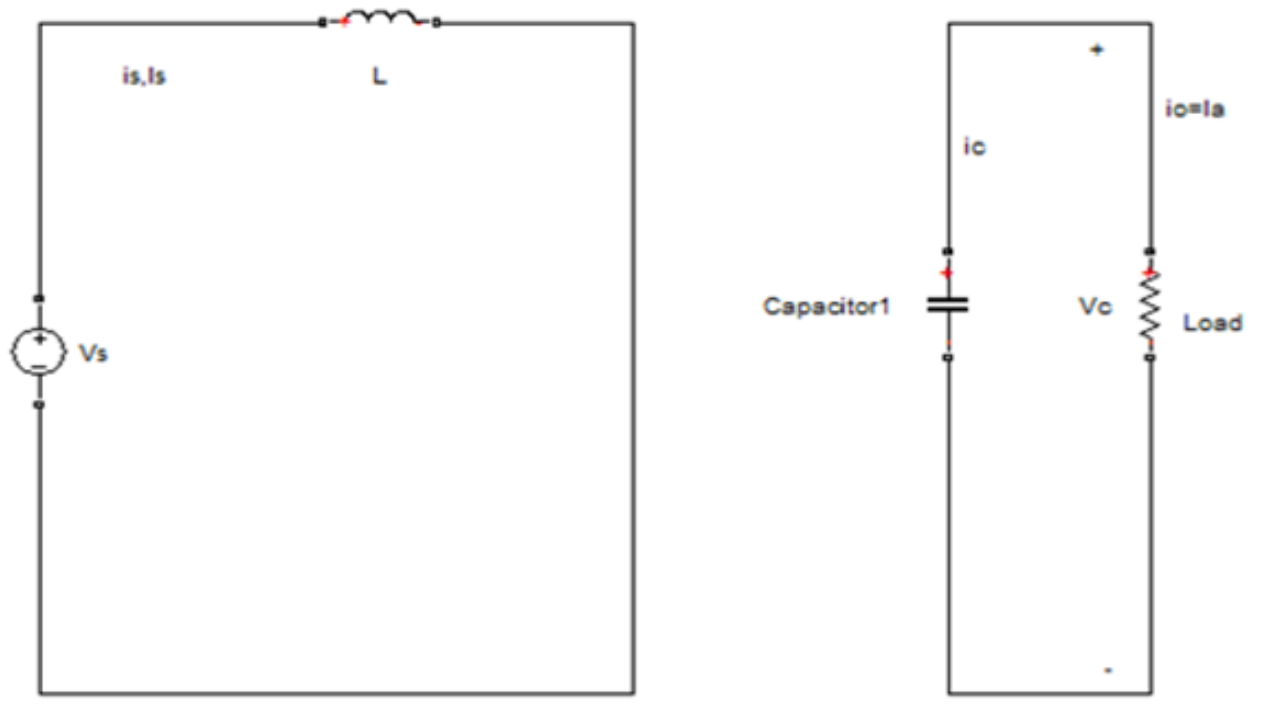

Figure 9: Boost converter operation of MODE I

Mode 1 begins when MOSFET is switched on at time t=0. The input current rises and flows through inductor L and MOSFET are shown in Figure 9.

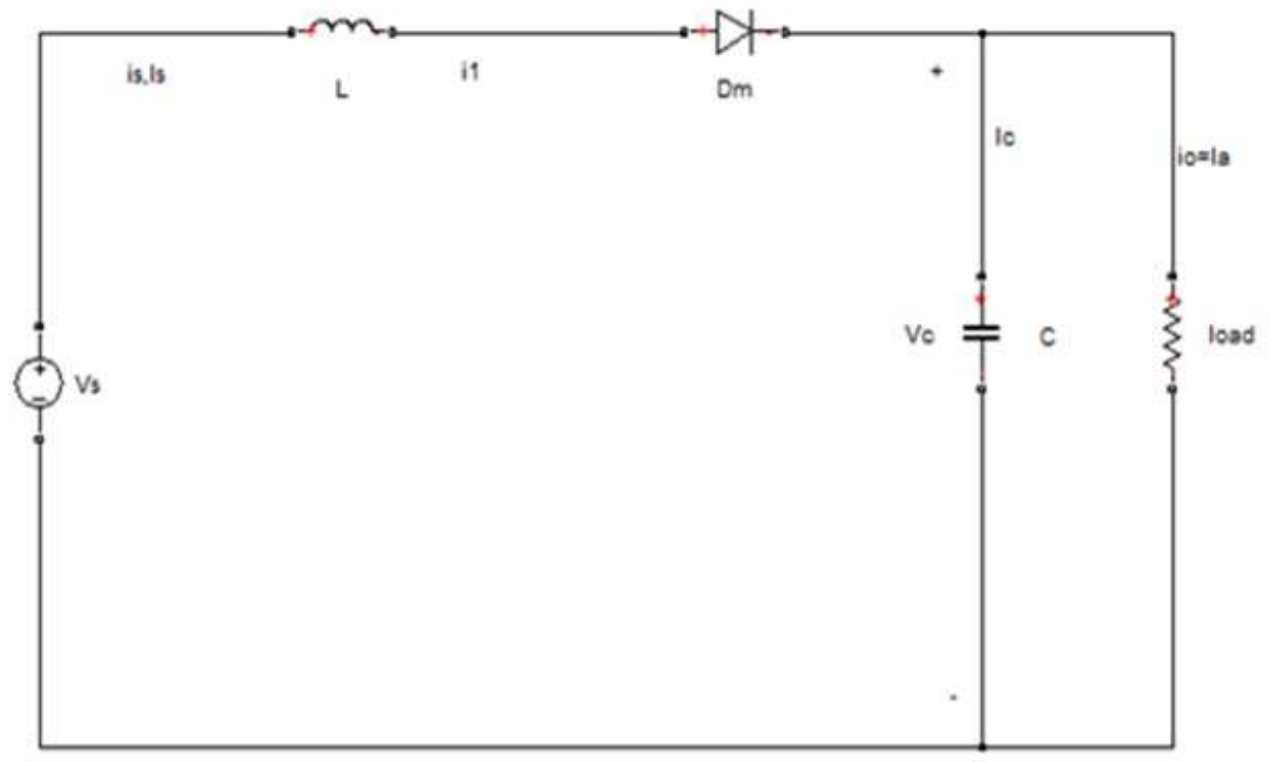

Figure 10: Boost converter operation of MODE II

Mode 2 begins when MOSFET is switched off at time t=t1. The input current now flows through L, C, load, and diode Dm. The inductor current falls until the next cycle. The energy stored in inductor L flows through the load is shown in Figure 10.The waveforms of the voltages and currents for boost converter are shown in Figure 11.

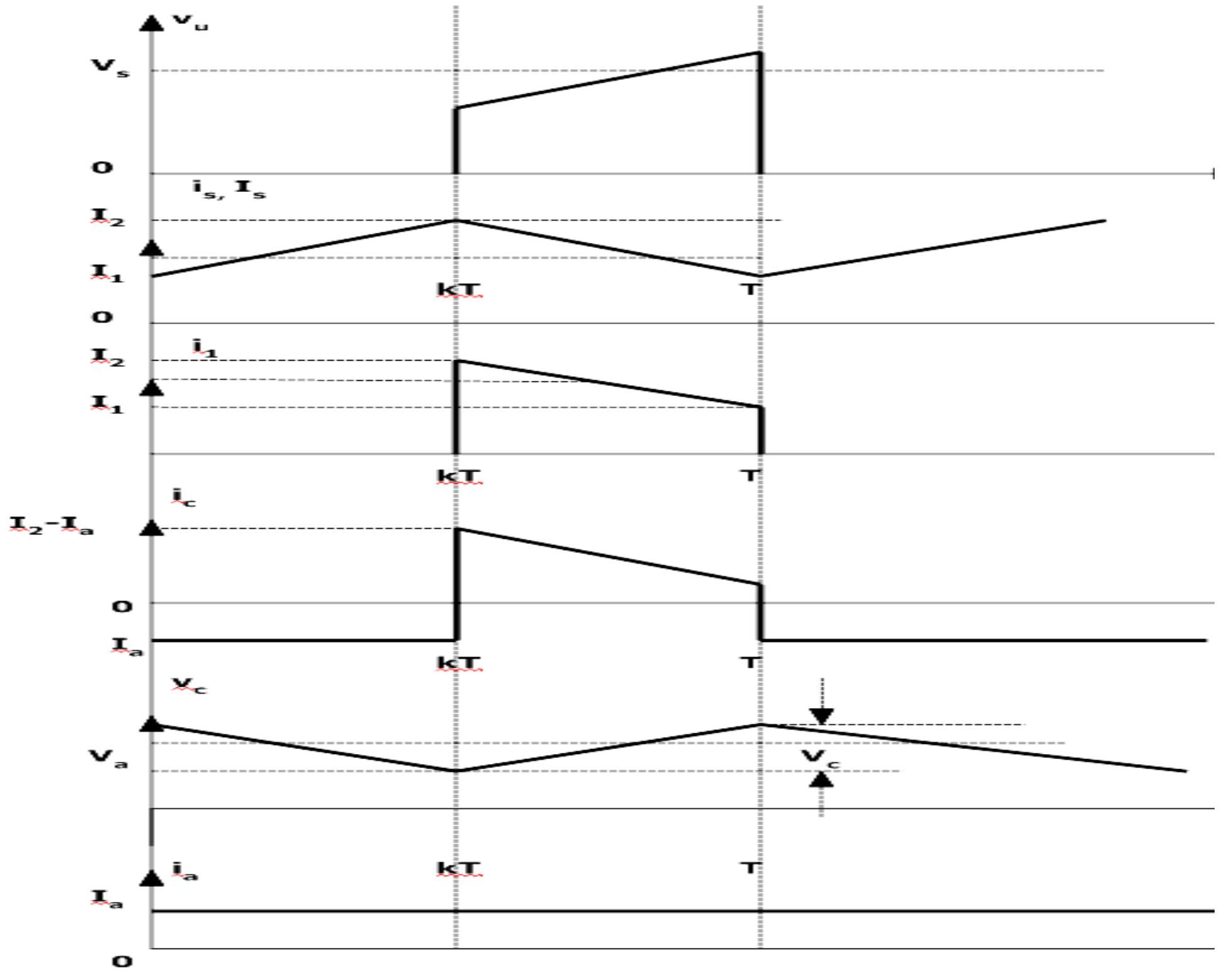

Figure 11: The waveforms of the voltages and currents for boost converter.

The voltage-current relation for the inductor L is:

$$i = \frac{1}{L}\int_0^t V dt + i_o \qquad \text{...... (2)}$$

$$V = L\frac{di}{dt}$$

For a constant rectangular pulse:

$$i = \frac{V_t}{L} + i_o$$

When the MOSFET is switched:

$$i_{pk} = \frac{(V_{in} - V_M)T_{on}}{L} + i_o$$

$$\Delta i = \frac{(V_{in} - V_M)T_{on}}{L} \qquad \text{....... (3)}$$

and when the MOSFET is switched off the current is:

$$i_o = i_{pk} - \frac{(V_{out} - V_{in} + V_D)T_{off}}{L}$$

$$\Delta i = \frac{(V_{out} - V_{in} + V_D)T_{off}}{L} \quad \ldots.(4)$$

Here $V_D$ is the voltage drop across the diode $D_{m,}$ and $V_M$ is the voltage drop across the MOSFET.

By equating through $\Delta i$, we can solve for $V_{out:}$

$$\frac{(V_{in} - V_M)T_{on}}{L} = \frac{(V_{out} - V_{in} + V_D)T_{off}}{L} \ldots\ldots.(5)$$

$$V_{in} - V_M D = (V_{out} + V_D)(1 - D)$$

$$V_{out} = \frac{(V_{in} - V_M D)}{(1 - D)} - V_D$$

Neglecting the voltage drops across the diode and the switch

$$V_{out} = \frac{V_{in}}{1 - D} \ldots\ldots.(6)$$

The active switch in the boost converter is a SiC MOSFET (1200V, 40A). A fast recovery diode is used as the freewheeling diode. The functioning principle of the boost is to excite the switch (SiC MOSFET) transistor with a duty cycle D produced by the MPPT control and when the switch is closed the inductor L is loading during T(D) time, afterwards the switch is opened, the inductor supplies the load through the diode during (1-D)T. For a DC-DC boost converter, the input–output voltage relationship for continuous conduction mode and design equation of L & C is given by:

$$c = \frac{I_o * D}{f_s * \Delta V_o} \ldots\ldots.(7)$$

$$L = \frac{V_s * D}{f_s * \Delta I_o} \ldots\ldots.(8)$$

Based on the design equations, the simulation parameters for SiC boost converter is shown in Table 2.

Table 1: Simulation parameters SiC DC-DC boost converter for PV

| Simulation parameters | Values |
|---|---|
| Photovoltaic panel | $V_{oc}$ = 31.1 V<br>$I_{sc}$ = 8.05 A<br>P = 250 W<br>N = 36 cells |
| SiC DC- DC Boost converter | Input= 31.1V<br>Output= 62.0V<br>C = 330 µF, 450 V<br>L=2mH, 15 A. |

**5. SiC MOSFET boost DC-DC converter powered by PV Source**

The proposed SiC converter is interfaced with PV and the PV cell is modeled using single diode model. A solar cell is the building block of a photovoltaic panel. A photovoltaic panel is developed by connecting many solar cells in series and parallel. A single photovoltaic cell can be modeled by utilizing current source, diode and two resistors as shown in Figure 12. It can be seen that the photovoltaic cell operates as a constant current source at low values of operating voltages and a constant voltage source at low values of operating current.

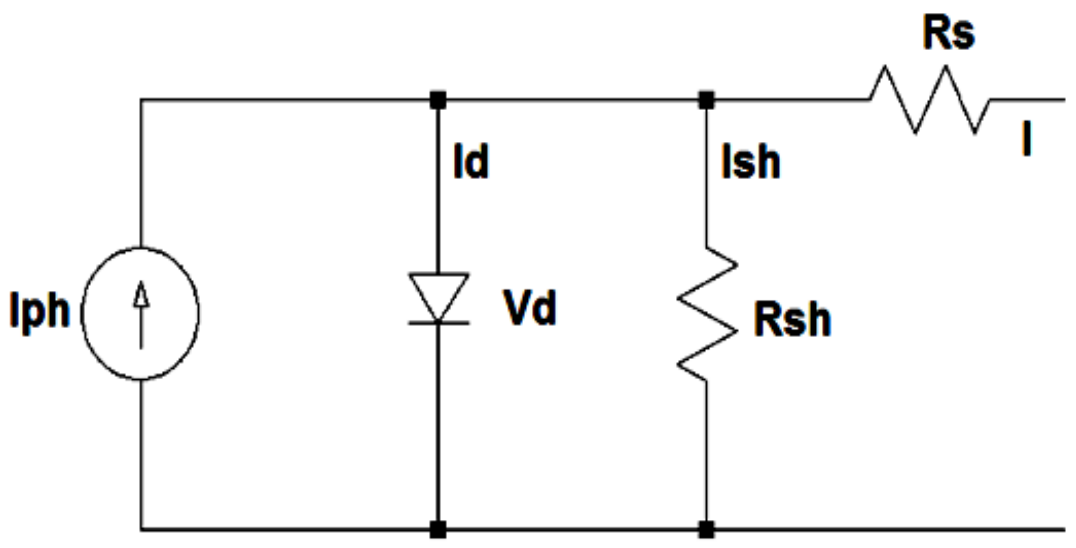

Figure 12: A single solar cell circuit model

The output efficiency of a photovoltaic cell is very low. In order to increase the output efficiency, different methods are to be analyzed to match the source and load side properly. So, to develop one such method is the Maximum Power Point Tracking control technique. This technique is used to obtain the maximum possible output power from a varying source power. In the photovoltaic systems, the V-I curve is non-linear, so it is difficult to get matched with the load. This technique is used for boost converter whose switching duty cycle is varied by using a Maximum power point tracking algorithm. A boost converter is used on the load side and a photovoltaic panel is used to power this converter. Various MPPT techniques are available in the literature, but this paper focuses on Incremental Conductance (INC) algorithm [5, 6].

**5.1. Incremental conductance method**

This method uses the photovoltaic panel incremental conductance dI/dV to compute the sign of dP/dV. When dI/dV is equal and opposite to the value of I/V (where dP/dV=0) the incremental conductance algorithm knows that the maximum power point tracking control is reached and thus it terminates and returns the corresponding value of operating voltage for maximum power point. Flow chart of incremental conductance method is shown in Figure 13. Moreover, this method tracks rapidly changing solar irradiation conditions more accurately than conventional method [7-10]. One complexity in this method is that it requires many sensors to operate and hence is economically less effective. Equations governing the proposed algorithm are as follows:

$$P = V * I \text{ ..... (9)}$$

$$\frac{dP}{dV} = \frac{d(V * I)}{dV} \text{ ....... (10)}$$

$$\frac{dP}{dV} = I * \left(\frac{dV}{dV}\right) + V * \left(\frac{dI}{dV}\right) \quad \text{ ....... (11)}$$

$$\frac{dP}{dV} = I + V * \left(\frac{dI}{dV}\right) \text{ ......... (12)}$$

When the MPPT is reached the slope

$$\frac{dP}{dV} = 0 \quad \text{ ....... (13)}$$

$$I + V * \left(\frac{dI}{dV}\right) = 0 \text{ ....... (14)}$$

$$\frac{dI}{dV} = -\frac{I}{V} \text{ ......... (15)}$$

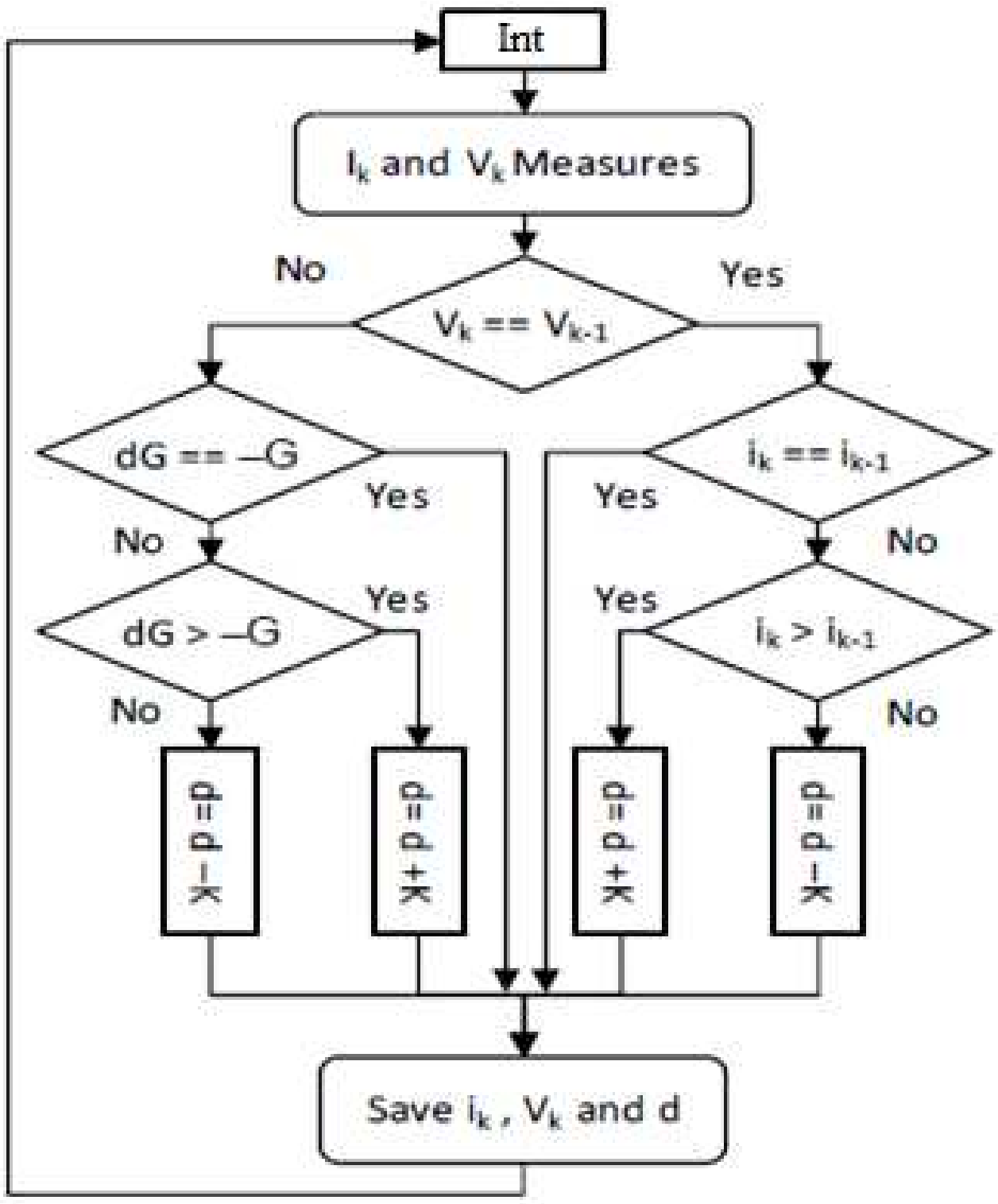

Figure 13: Flowchart of Incremental conductance method.

Simulink model of the IC based MPPT control of SiC DC– DC boost converter based photovoltaic power generation is shown in Figure.14.

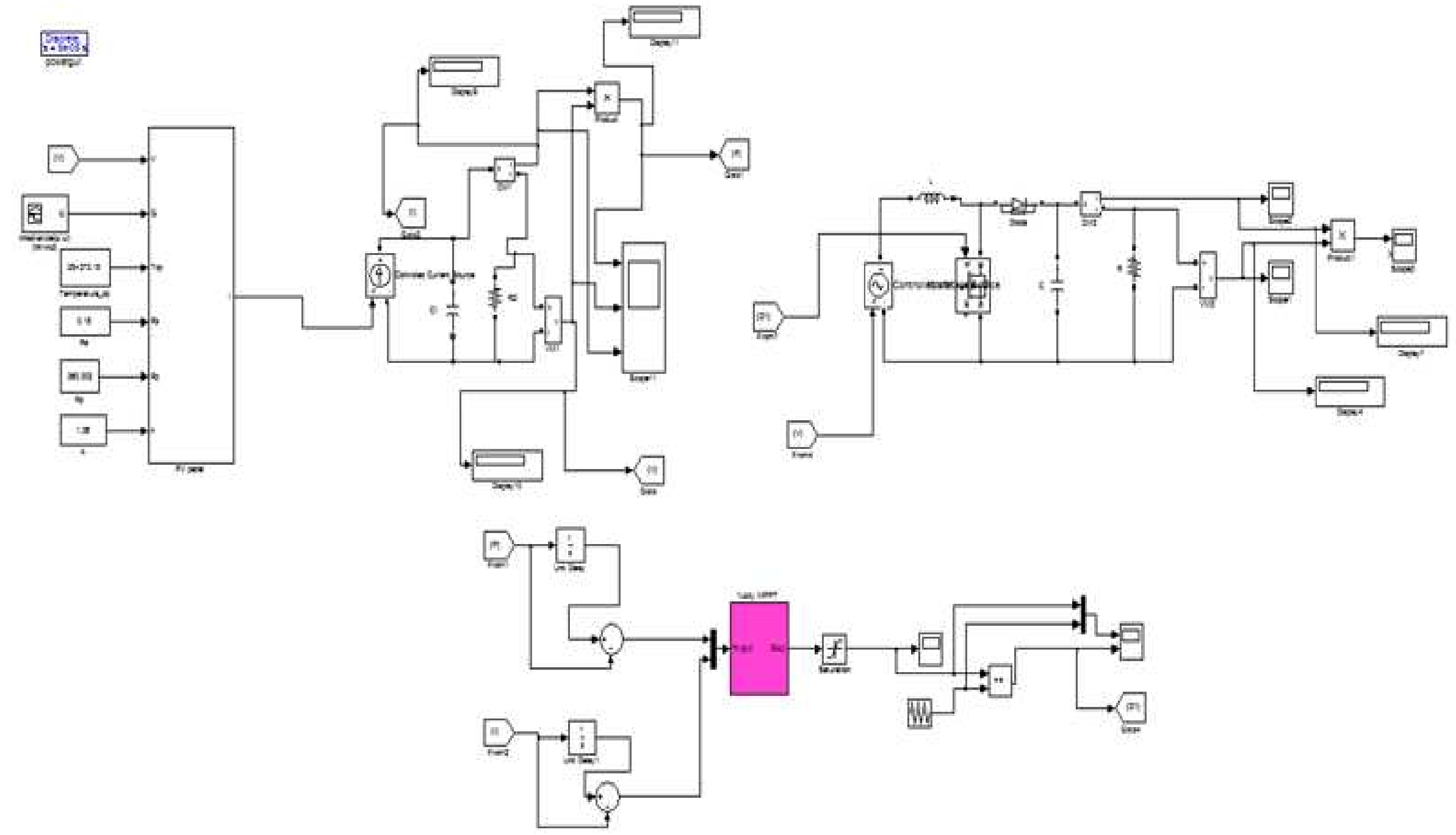

Figure 14: Simulink model of the IC MPPT control of SiC DC – DC boost converter based photovoltaic power generation.

The gate pulse of SiC MOSFET switch with 50 % duty cycle is shown in Figure.15 which is obtained by PWM control.

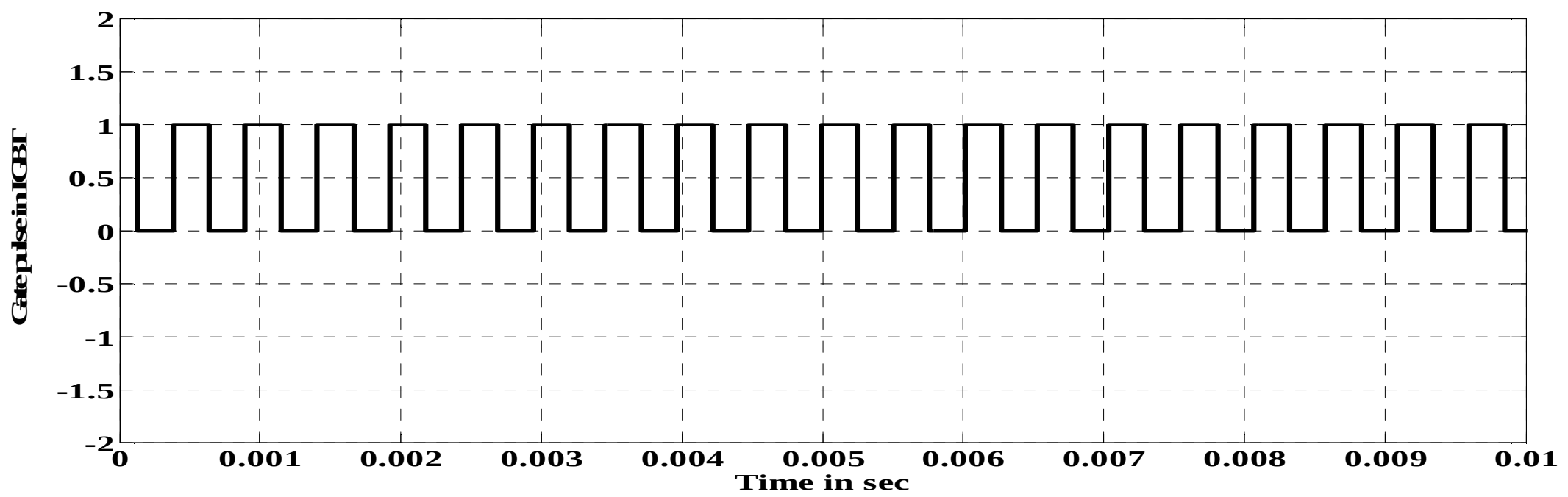

Figure .15: Gate pulse of SiC MOSFET

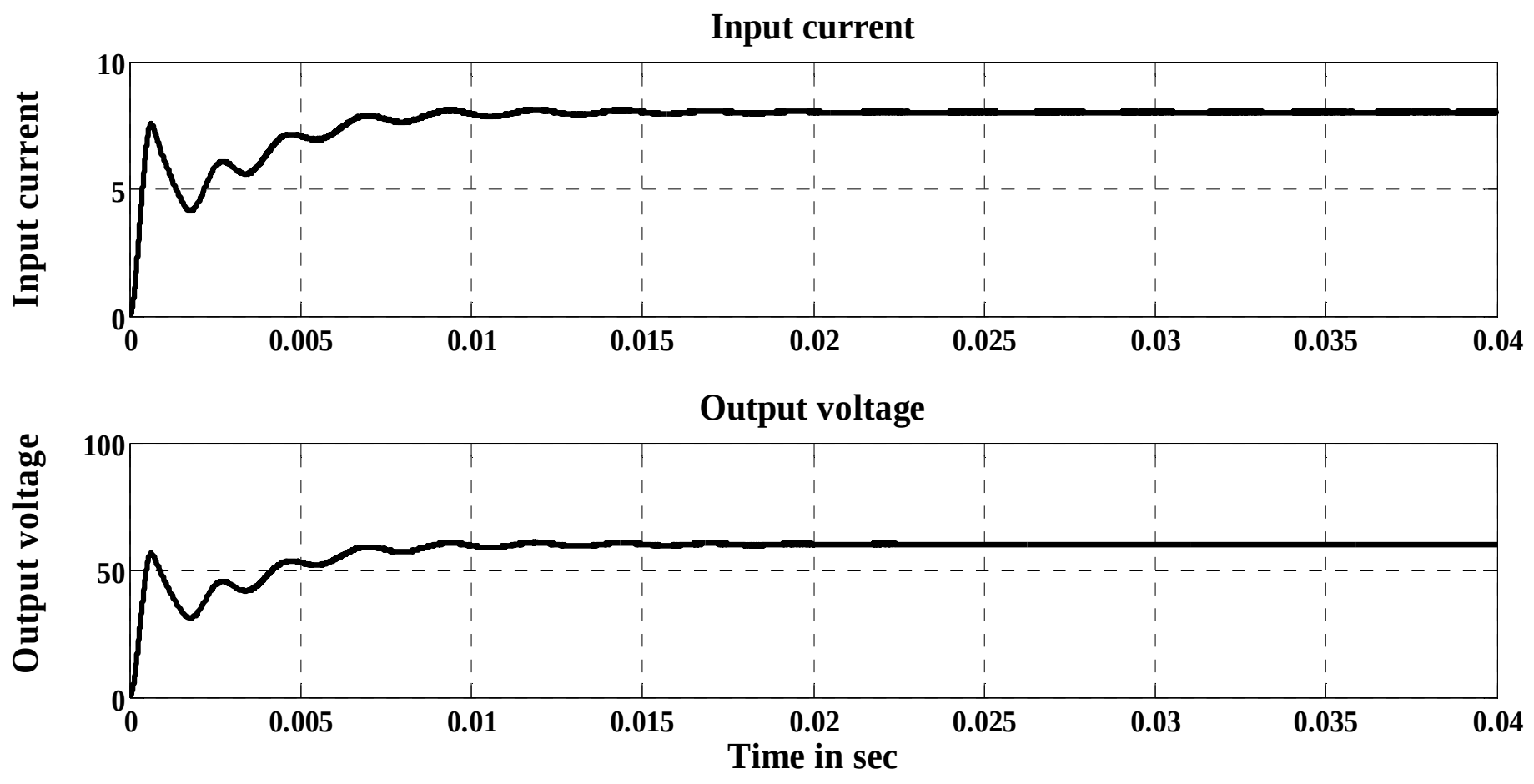

Figure .16: Input current and Output voltage of boost converter.

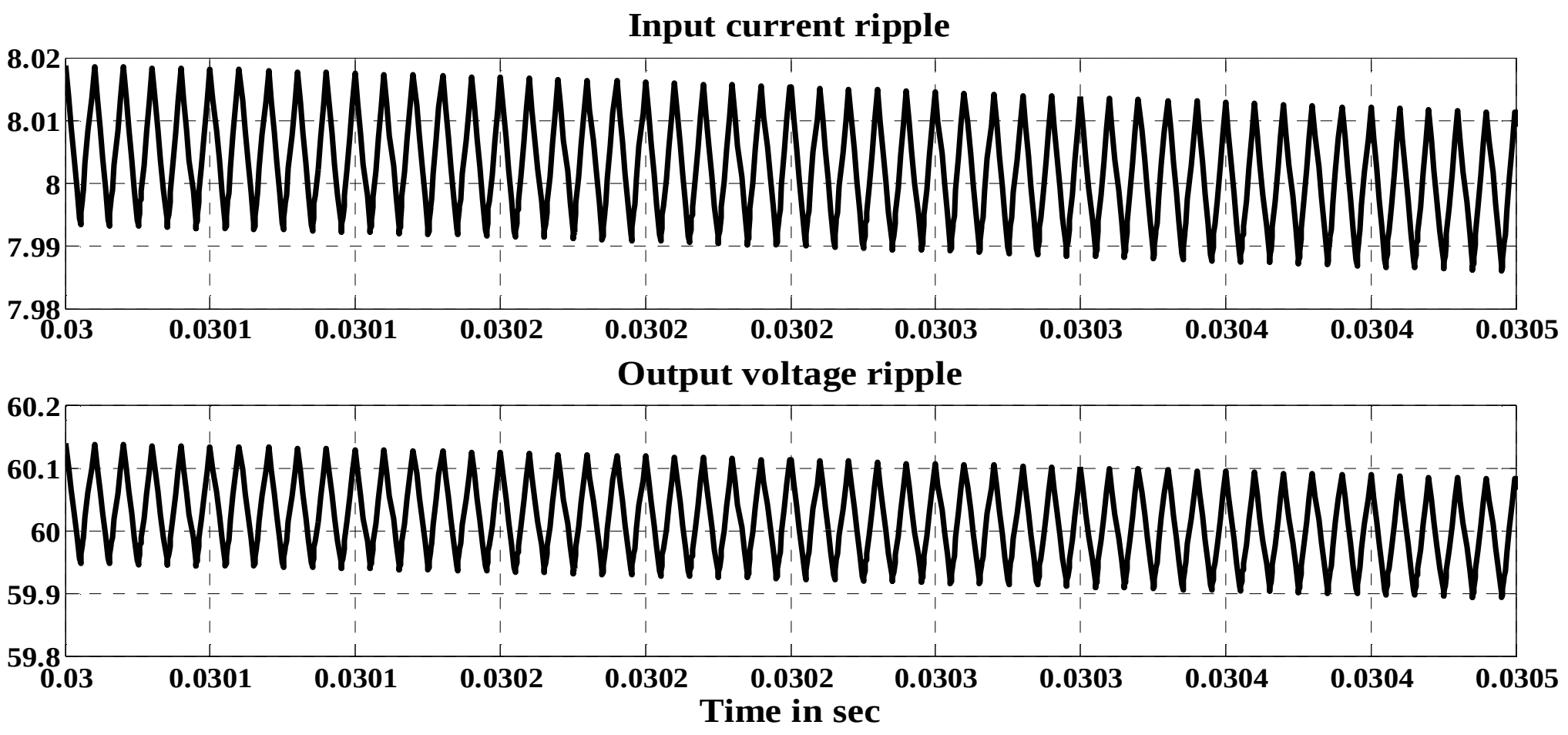

Figure .16: Input current ripple and Output voltage ripple of boost converter.

The DC- DC boost converter output voltage is about 60 Volts and input current of DC-DC boost converter is about 8 A as shown in Figures.16& 17.

## 6. EXPERIMENTAL SETUP OF PHOTOVOLTAIC POWER GENERATION

The hardware set-up for the SiC MOSFET boost converter for photovoltaic power generation is shown in Figure.18. Photovoltaic panel and SiC boost converter specifications are shown in table 3.

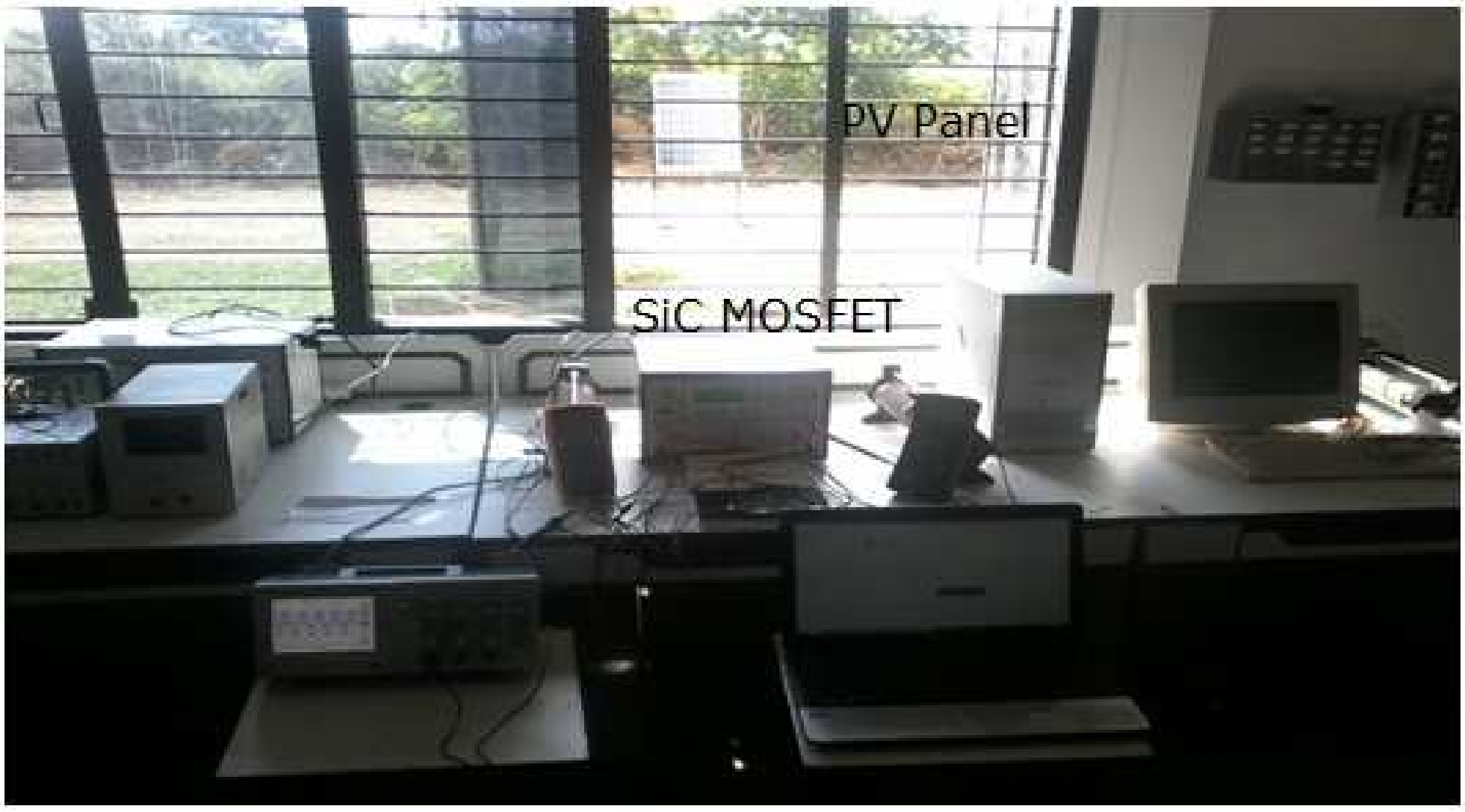

Figure.18.Hardware set-up for SiC MOSFET Boost converter for PV

Table 2: Specifications of PV Panel & SiC Boost Converter

| **Parameters** | **Values** |
|---|---|
| $V_{oc}$ | 31.1 V |
| $I_{sc}$ | 8.05 A |
| $P_{max}$ | 250 W |
| Insolation W/m$^2$ | 1000W/m$^2$ |
| System efficiency | 76.72 W |
| SiC MOSFET | SCT 2080KE, 1200 V,40 A,1KW |
| Fast recovery diode | MUR 3060,600 V, 30A |
| Input Capacitance | C1 = 470 μF, 50 V |
| Output Capacitance | C2 = 330 μF, 450 V |
| Inductance | L1=2mH, 15 A. |
| Switching Frequency | 150kHz |

The experimental P-V and V-I characteristics are shown in Figure.19.

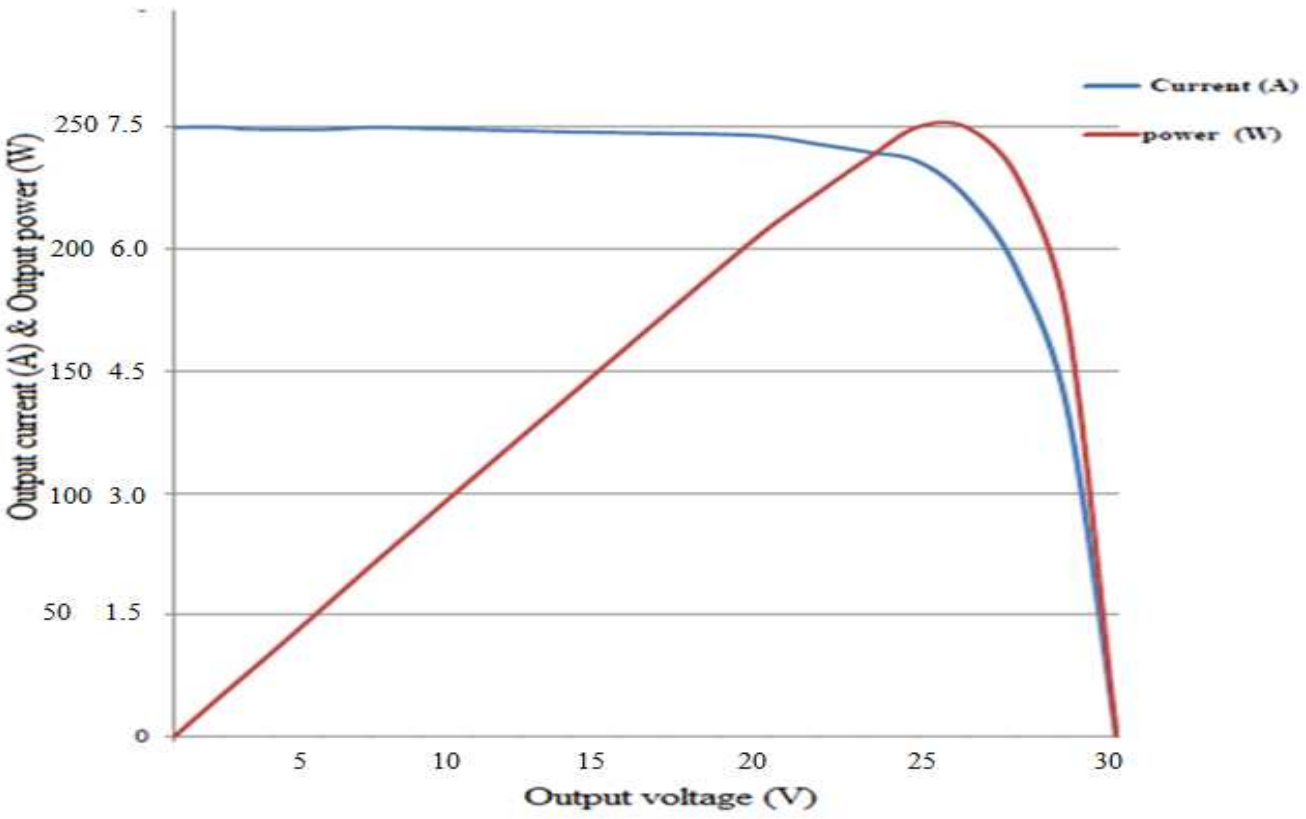

Figure 19: P-V & V-I characteristics of a photovoltaic panel

## 6.1 Implementing IC based MPPT on FPGA

The IC based MPPT is implemented on the FPGA board. Then, the DC-to-DC is hooked up and connected to FPGA [11-14]. Figure.20 shows the logic circuit diagram in Xilinx ISE 14.1 software for the IC based MPPT and other components. The output of the controller is connected with a PWM module designed on the FPGA. The experimental PWM frequency of the modulating signal is about 50 KHz. The output of the PWM is examined using DSO by changing the values of the MPPT as shown in Figure. 21.

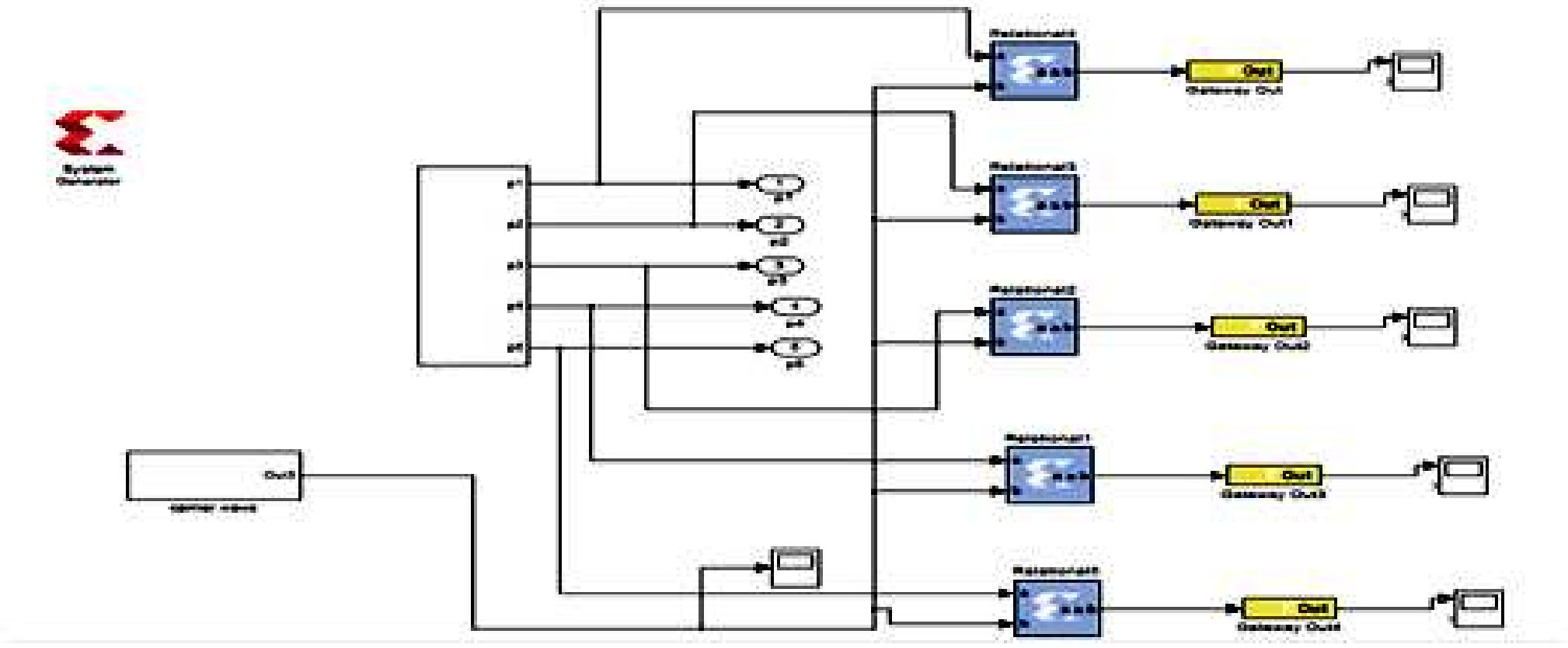

Figure 20: The logic circuit diagram in Xilinx software for the MPPT.

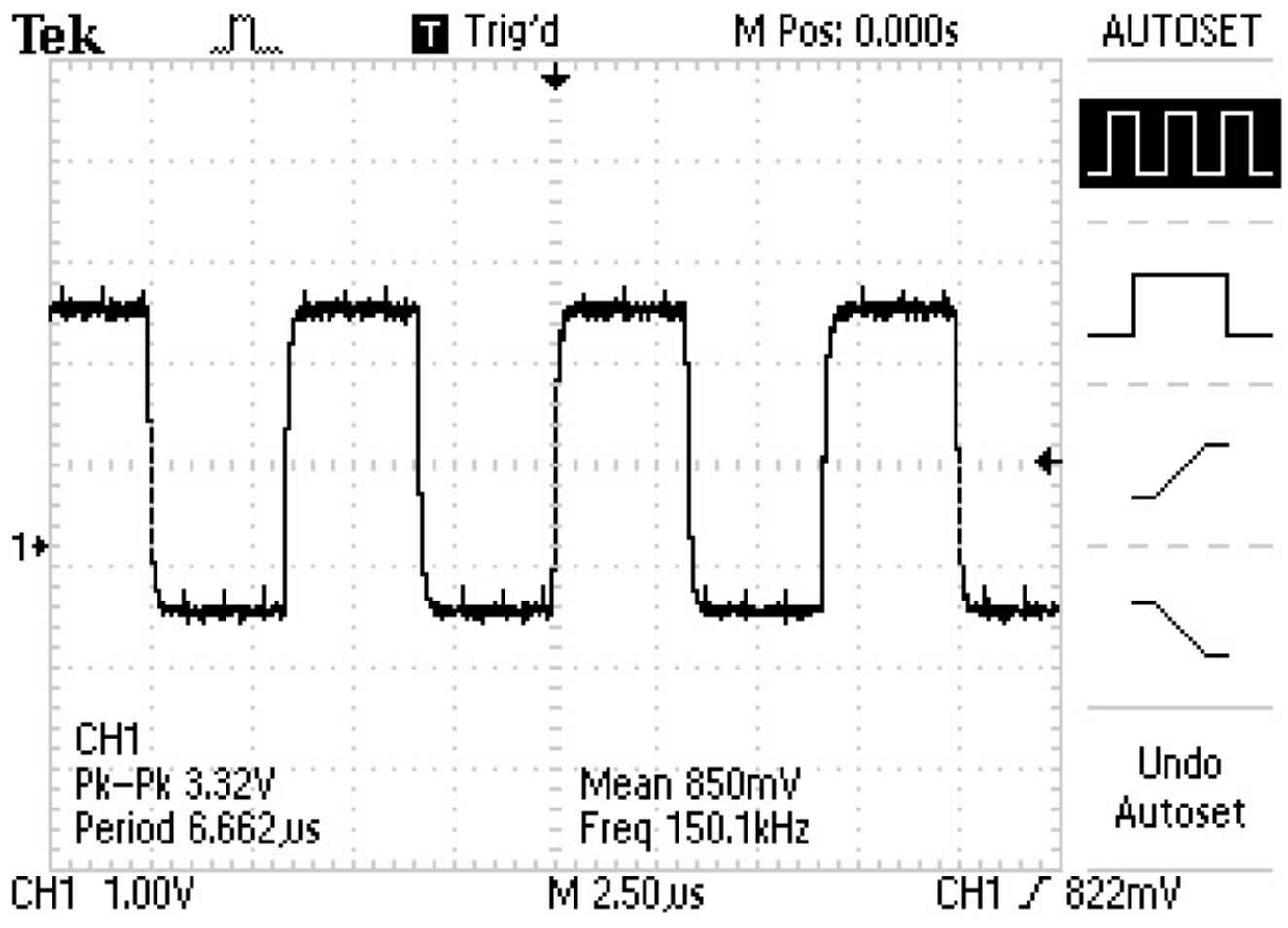

Figure.21.The change in the duty cycle of the PWM output (50% duty cycle)

The input to the converter is about 28.2V and output voltage of DC-DC boost converter is about 62.4 V as shown in Figure.22. The SiC MOSFET switches at 50 % duty cycle.

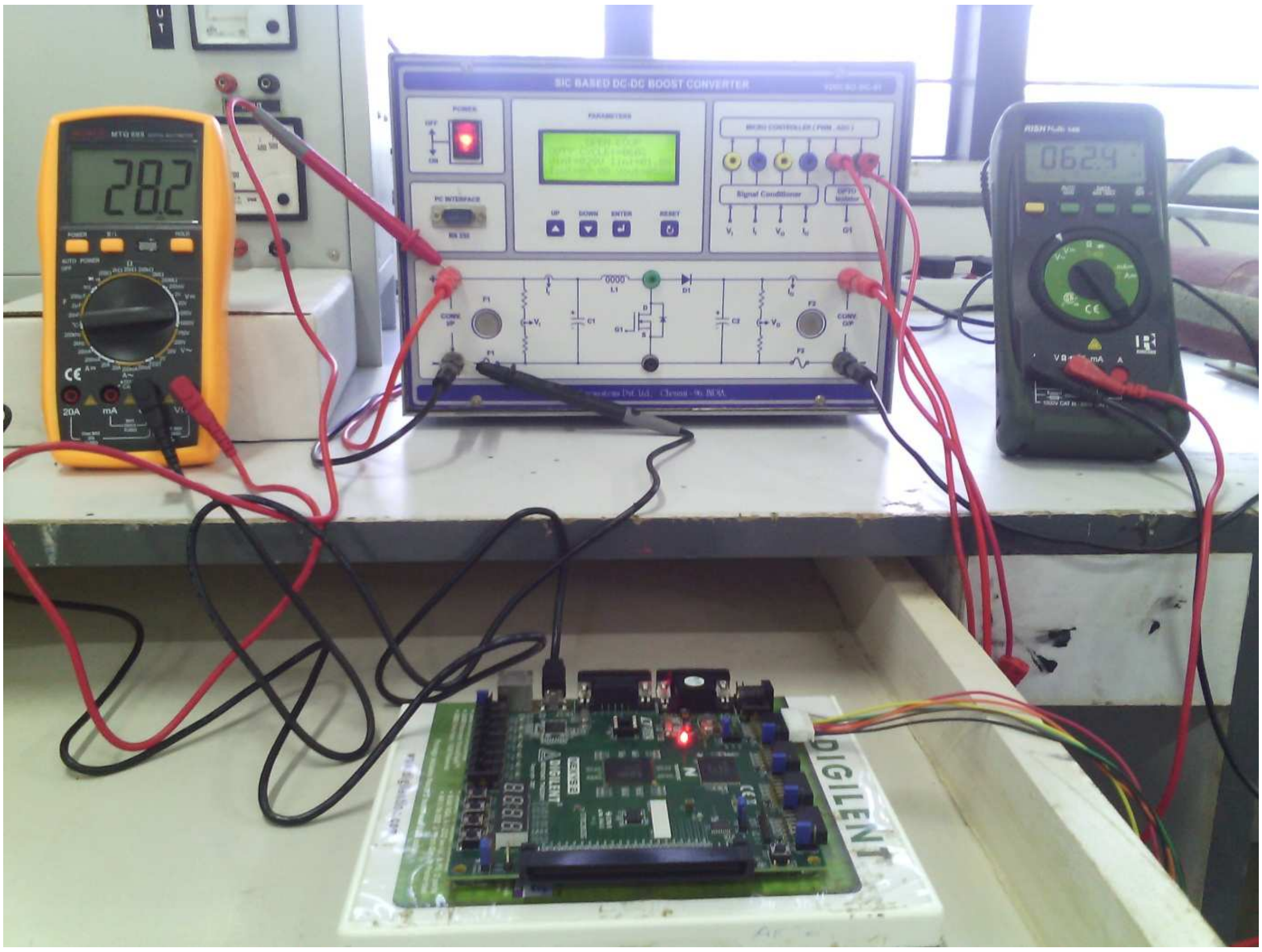

Figure 22.Input and Output voltage of step up boost converter using MPPT.

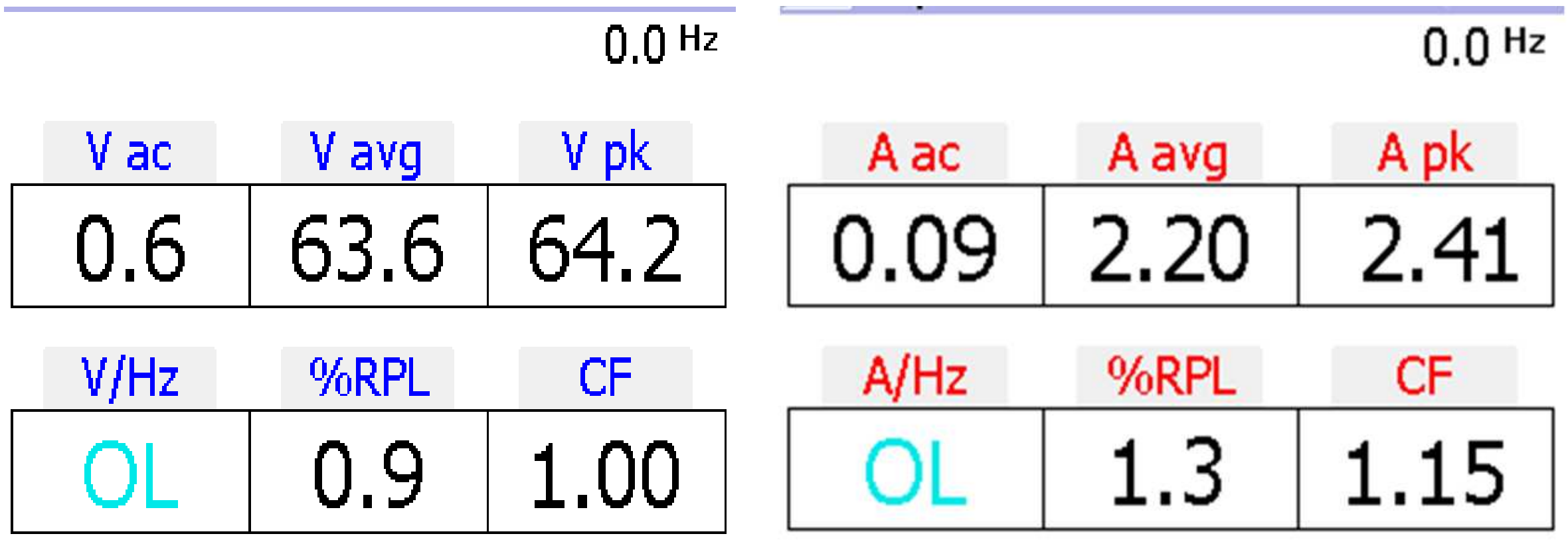

Figure 23.Output voltage ripple and input current ripple of SiC DC-DC boost converter for 50% duty cycle.

Output voltage ripple and input current ripple of SiC DC-DC boost converter measured using PQ analyzer is about 0.9 % and 1.3 % as shown in Figure.23.

## 7.Performance analysis of the proposed SiC Boost Converter for PV

The performance parameters of the proposed SiC boost converter such as conduction, switching losses, thermal analysis and ripple calculations for output voltage and input current are computed and it is compared with the conventional Si MOSFET. The parameters are explained as follows:

### 7.1 Loss Calculation for the proposed SiC MOSFET based boost converter

The power electronic devices dissipate power when they are used in various applications. If they dissipate too much power, the devices can fail and it may also damage the other system components. Power dissipation in power electronics devices is a common thing. So the converter designer must understand how to minimize the power dissipation in the power electronics devices. Table 4 shows conduction and switching loss calculation for SiC MOSFET which is obtained based on Eqs. (16-18).

$$P_{MOSFET} = P_{Switching} + P_{Conduction} \quad \text{.........(16)}$$

$$P_{Conduction} = I_{on}^2 * R_{DS(ON)} \quad \text{.........(17)}$$

$$P_{SW} = \frac{t_{sw,on} * V_{off} * I_{on} * f_{sw}}{2} + \frac{t_{sw,off} * V_{off} * I_{on} * f_{sw}}{2} \quad \text{.........(18)}$$

The SiC MOSFET under transient and saturation region and its corresponding dynamic parameters are shown in Table 1.From Table 1, it is inferred that the on-time and off time for SiC MOSFET is greatly reduced compared to conventional Si MOSFET.

**Table 3**: Dynamic parameters

| Parameter | SiC MOSFET (SCT 2080 KE) | Si MOSFET (APT29F100L) |
|---|---|---|
| Turn on time ($t_{on}$) | 35 ns | 39 ns |
| Turn off time ($t_{off}$) | 76 ns | 130 ns |
| Fall time ($t_f$) | 36 ns | 35 ns |
| Rise time ($t_r$) | 22 ns | 33 ns |

**Table 4**: Comparison of loss calculation for SiC and Si MOSFET

| Loss calculation | SiC MOSFET | Si MOSFET | Fast recovery Diode | Boost Diode |
|---|---|---|---|---|
| **Conduction loss** | 5.12 W | 9.58 W | 0.454 W | 1.25 W |
| **Switching loss** | 1.6 W | 2.3 W | 0.016 W | 0.525w |
| **Total losses** | 6.72 W | 11.88 W | 0.47W | 1.77 W |

Table 4 shows that the total losses are reduced in SiC MOSFET compared to Si MOSFET and thereby, the efficiency of the SiC based boost converter is improved .

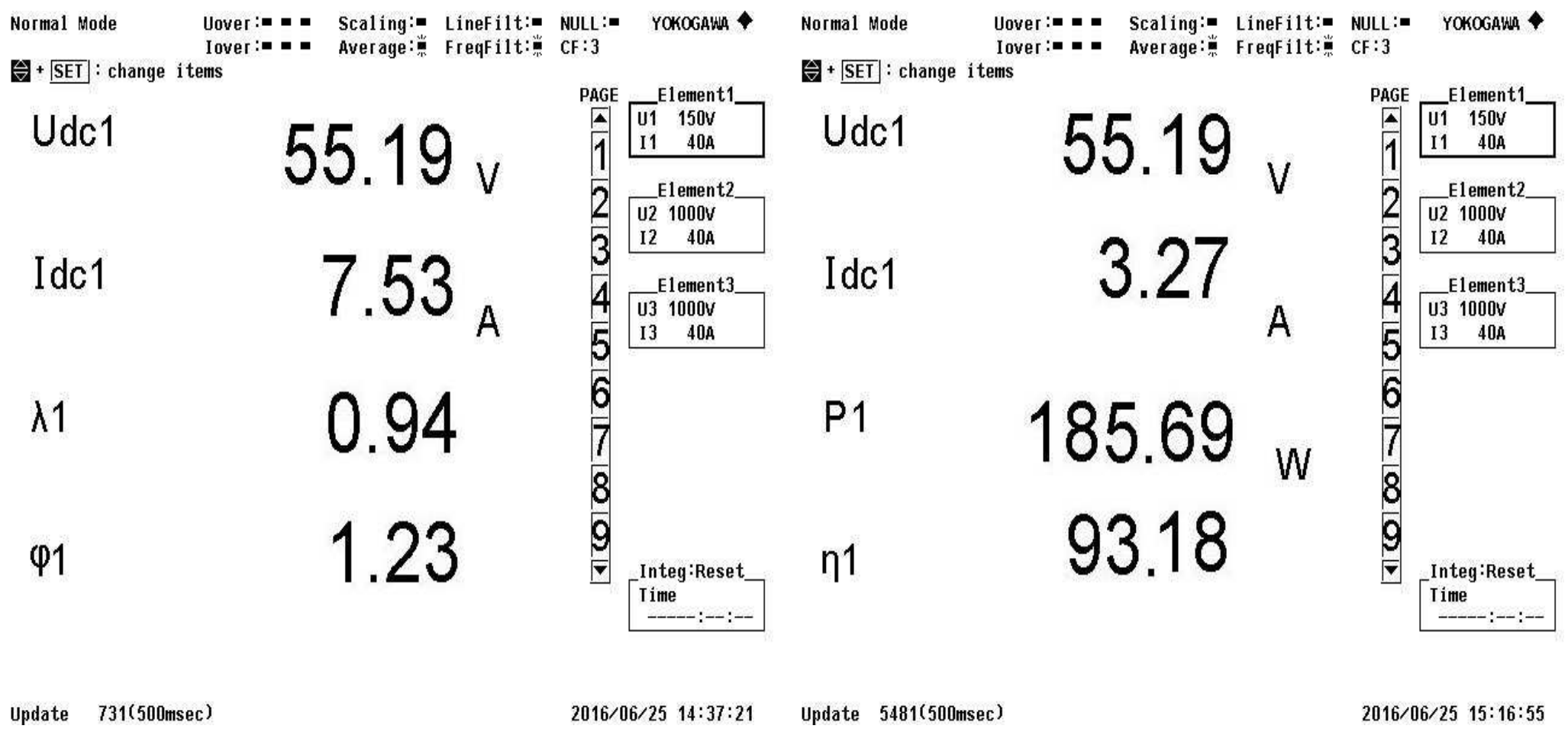

**Figure 24 Performance analysis of SiC DC-DC boost converter with battery charging (48 V,100 Ah) for 50% duty cycle**

The simulation results shows that the output voltage ripple and input current ripple of boost converter is about 0.98V & 1.3A. Output voltage ripple ($\lambda_1$),input current ripple ($\varphi_1$) and efficiency ($\eta$1)of SiC DC-DC boost converter measured using PQ analyzer is about 0.94V, 1.3A and 93.18 % as shown in Figure 24. The simulation results verified experimentally.

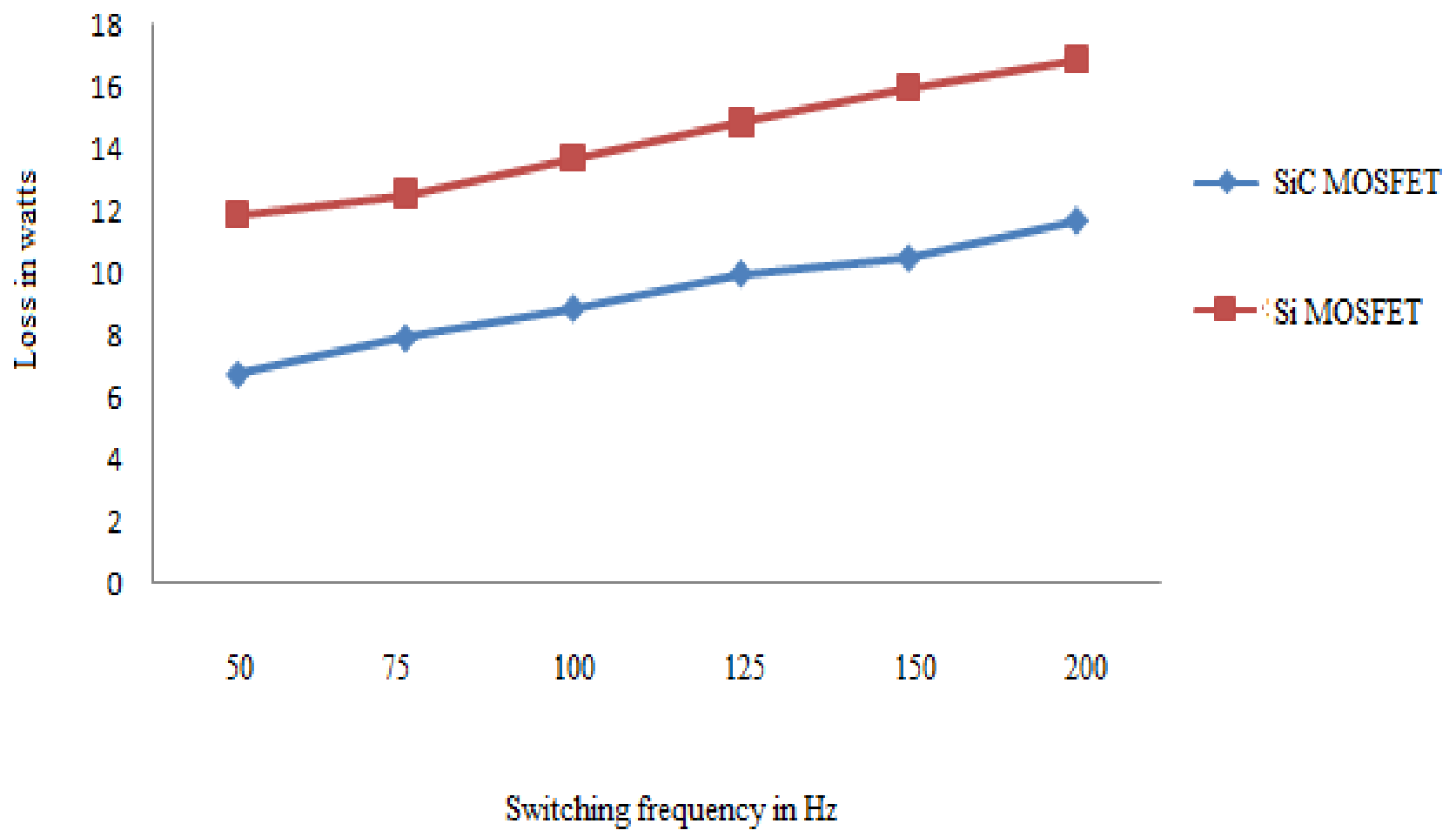

Figure 25: Loss Vs Switching Frequency

Figure 25 shows that the comparison between Si MOSFET and SiC MOSFET for different switching frequency and SiC shows a reduced loss as frequency increases compared to Si.

**Table 5:** Comparison between Si MOSFET and SiC MOSFET

| **Duty Cycle %** | **SiC DC-DC Output voltage ripple %** | **Si DC-DC Output voltage ripple %** | **SiC DC-DC Input current ripple %** | **Si DC-DC Input current ripple %** |
|---|---|---|---|---|
| 40 | 1.1 | 3.2 | 1.54 | 5.45 |
| 50 | 0.9 | 2.9 | 1.3 | 5.38 |
| 60 | 0.83 | 2.8 | 1.25 | 5.26 |
| 70 | 0.77 | 2.8 | 1.17 | 5.22 |

Table 5 shows the comparison between Si MOSFET and SiC MOSFET. By employing SiC MOSFET, it was found that the output voltage ripple and input current ripple are reduced compared to conventional MOSFET.

**8. Conclusion**

This paper has proposed a SiC MOSFET based boost converter for PV applications. To extract maximum power from PV, an incremental conductance MPPT was implemented on FPGA Spartan-3E board for PV system. It is observed that the proposed SiC converter gives a reduced output voltage(0.93V), input current ripple(0.13A) and efficiency(93%). Moreover, with SiC, the conduction and switching losses are greatly reduced and therefore the SiC MOSFET based boost converter is a suitable choice for solar PV pre-regulator.